\documentclass[12pt,hidelinks]{article}

\RequirePackage{amsthm,amsmath,amsfonts,amssymb}
\RequirePackage[numbers]{natbib}
\RequirePackage[colorlinks,citecolor=blue,urlcolor=blue,backref=page,backref=page]{hyperref}
\RequirePackage{graphicx}
\RequirePackage{tikz}
\usetikzlibrary{fit, positioning}

\usepackage{caption}
\usepackage{subcaption}
\usepackage{booktabs}
\usepackage{multirow}
\usepackage{bm}
\usepackage{bbm}
\usepackage{soul}
\usepackage{authblk}

\theoremstyle{plain}

\newtheorem{prop}{Proposition}

\theoremstyle{definition}

\theoremstyle{remark}

\newcommand\sdots{\hbox to 0.5em{.\hss.\hss.}}

\newcommand{\keywords}[1]
{
  \small	
  \textbf{\textit{Keywords---}} #1
}

\usepackage{mathtools} % Bonus
\DeclarePairedDelimiter\norm\lVert\rVert
    
\title{Bayesian Time-Varying Tensor Vector Autoregressive Models for Dynamic  Effective Connectivity}
	
\author[1]{Wei Zhang\footnote{{wei.zhang@phd.unibocconi.it}}}
\author[2,3]{Ivor Cribben\footnote{{cribben@ualberta.ca}}}
\author[1]{Sonia Petrone\footnote{{sonia.petrone@unibocconi.it}}}
\author[4]{Michele Guindani\footnote{{michele.guindani@uci.edu}}}

\affil[1]{Department of Decision Sciences, Bocconi University}
\affil[2]{Department of Accounting and Business Analytics, Alberta School of Business}
\affil[3]{Neuroscience and Mental Health Institute, University of Alberta}
\affil[4]{Department of Biostatistics, University of California, Los Angeles}

\begin{document}
\maketitle

\begin{abstract}
In contemporary neuroscience, a key area of interest is dynamic effective connectivity, which is crucial for understanding the dynamic interactions and causal relationships between different brain regions. Dynamic effective connectivity can provide insights into how brain network interactions are altered in neurological disorders such as dyslexia. Time-varying vector autoregressive (TV-VAR) models have been  employed to draw inferences for this purpose. However, their significant computational requirements pose challenges, since  the  number  of  parameters  to  be  estimated  increases quadratically  with  the  number of  time series. In this paper, we propose a computationally efficient Bayesian time-varying  VAR approach. For dealing with large-dimensional  time  series, the proposed framework employs a tensor decomposition for the VAR coefficient matrices at different lags.  
Dynamically varying connectivity patterns are captured by assuming that at any given time only a subset of components in the tensor decomposition is active. Latent binary time series select the active components at each time via an innovative and parsimonious Ising model in the time-domain. Furthermore, we propose parsity-inducing priors to achieve  global-local shrinkage of the VAR coefficients, determine automatically the rank of the tensor decomposition and guide the selection of the lags of the auto-regression. We show the performances of our model formulation via simulation studies and data from a real fMRI study involving a book reading experiment.  % We find that changes in the effective connectivity coincide with events in the book and the connectivity hubs identified are consistent with the reading literature.
\end{abstract}

\keywords{fMRI experiments, Tensor factorization, Granger Causality, Shrinkage priors, Temporal Ising model, $\text{MC}^3$ algorithm}

\section{Introduction}
\label{Introduction}

Investigating the integration among different brain regions to understand how cognitive information is processed and distributed is a central goal of many functional magnetic resonance imaging (fMRI) experiments. 
Neuroscientists typically distinguish between {\it functional connectivity}, which measures the undirected associations, or temporal correlation, between the fMRI time series observed at different locations, and {\it effective connectivity}, which estimates the directed influences that one brain region exerts onto other regions \citep{friston2011functional, zhang2015, DuranteGuindani2020}.  Unlike functional connectivity, which only measures the correlation between brain regions, effective connectivity aims to determine the direction and influence one brain region exerts over another. 
This offers researchers a way to explore causal relationships in brain activity.
Effective connectivity can also provide insights into how brain network interactions are altered in neurological and psychiatric disorders, aiding in the development of diagnostic markers and improving our understanding of disease mechanisms.  By identifying normal and abnormal effective connectivity patterns, researchers can develop targeted therapies.  For example, for reading disorders such as dyslexia \citep{fiecas,finn}, understanding altered brain networks can help in designing more effective intervention strategies, including neurostimulation or behavioral therapies.  The study of effective connectivity can also shed light on how learning and experience shape brain network interactions over time, contributing to our understanding of neural plasticity. 

One way to model effective connectivity is via a vector auto-regression (VAR) model, a widely-employed framework for estimating  temporal (Granger) casual dependence in fMRI experiments \citep[see, e.g.][]{gorrostieta2013hierarchical,  chiang2017bayesian}. 
 Furthermore, it is now recognized that the connectivity patterns may vary dynamically throughout the fMRI experiment, as we anticipate in our motivating dataset \citep{ZARGHAMI2020}.
This is supported by recent studies in neuroscience that emphasize the importance of describing changes in brain connectivity in response to stimuli in task-based settings \citep{cribben, ofori, anastasiou} and due to spontaneous fluctuations in resting-state fMRI \citep{hutchison2013dynamic, taghia2017bayesian, PARK2018, warnick2018bayesian, ZARGHAMI2020}. A Markov-switching VAR model formulation has been utilized by \citet{samdin2016unified} and \citet{ombao2018statistical} to characterize dynamic connectivity regimes among a few selected EEG channels. \citet{li2020bayesian} developed a stochastic block-model state-space multivariate autoregression to investigate how abnormal neuronal activities start from a seizure onset zone and propagate to otherwise healthy regions using intracranial EEG data. 
 
VAR models can become computationally intensive when analyzing large-dimensional time-series,  since the number of parameters to estimate increases quadratically with the  number of time-series, often exceeding the number of observed time points. Thus, several approaches have been proposed to enforce sparsity of the VAR coefficient matrix, either by using penalized-likelihood methods \citep{Shojaie:2010aa,basu2015} or - in a Bayesian setting  - by using several types of shrinkage priors \citep{primiceri2005time, koop2013forecasting,giannone2015prior}. Alternatively, dimension reduction techniques have been employed to reveal and exploit a lower dimensional structure embedded in the parameter space. For example, \citet{velu1986reduced} decompose the VAR coefficient matrix as the product of lower-rank matrices. \citet{wang2021high} have proposed an $L_1$- penalized-likelihood approach where a tensor decomposition is employed to represent the elements of the VAR coefficient matrices. \citet{luo2022bayesian} propose a static Bayesian VAR that decomposes the coefficient matrix into a low-rank tensor where a multiplicative gamma prior and an adaptive inferential scheme are used to estimate the tensor margins and rank parameters. Similarly, \citet{fan2022bayesian} developed a Bayesian random effects VAR model for large-dimensional multi-subject panel neuroimaging data where the dimension reduction is achieved by applying a Tucker tensor decomposition. \citet{billio2023dynamic} propose a
Bayesian autoregressive tensor framework to model the (static) parameters of a time-series regression, allowing for both tensor-valued multi-layer responses and covariates, and employing impulse response analysis to study the propagation of shocks across both the network and the layers. 

In this paper,  motivated by an experimental study on dynamic effective connectivity patterns arising when reading complex texts, we propose a computationally efficient time-varying Bayesian VAR approach for modeling large-dimensional time series. Similarly as in \citet{wang2021high}, we assume a tensor decomposition for the VAR coefficient matrices at different lags.  A novel feature of the proposed approach is that  we capture dynamically varying connectivity patterns by assuming that -- at any given time -- the VAR coefficient matrices are obtained as a sum of a subset of active components in the tensor decomposition. This sum representation relies on latent indicators of brain activity, that we model through an innovative use of an Ising model on the time-domain, to select what components are active at each time. With respect to Hidden Markov Models -- typically employed in the fMRI literature to capture   transitions across brain states dynamically over time -- the Ising specification models the time-varying activations as a function of only two parameters. The resulting binary time series still maintains a Markovian dependence, but the  Ising model naturally assigns  a higher probability mass to  non-active (zeroed) components to encourage sparsity of the representation and it favors similar selections at two consecutive time points, reflecting the prior belief that the coefficients are changing slowly over time. Furthermore, we show that the Ising model can be represented as the joint distribution of a so-called (new) discrete autoregressive moving average (NDARMA) model \citep{jacobs1983stationary}, a result which is helpful for prior elicitation.  \\
The remaining components of the model are designed to encourage sparsity in the tensor structure and to ascertain model complexity directly from the data through the posterior distribution. Specifically, we employ a multi-way Dirichlet generalized double Pareto prior \citep{guhaniyogi2017bayesian, billio2023dynamic}, which allows for global-local shrinkage of the VAR coefficients. This choice enables the model to implicitly determine the effective rank of the tensor decomposition based on the observed data. An additional feature of our approach is that it assumes an \emph{increasing}-shrinkage prior \citep{legramanti2020bayesian} to guide the selection of the lags of the auto-regression, eliminating the need to rank models of different order based on model selection information criteria. We show how inference on the VAR coefficient matrices can be used to create time-varying Granger causality networks, enabling a detailed analysis of dynamic connections over time.  In our fMRI application involving a book reading experiment, we find that changes in the effective connectivity coincide with key events in the book and the connectivity hubs identified are consistent with the reading literature.  Although only controls were studied in this data, the new methodology could be used in a study of neurological and psychiatric disorders (such as dyslexia) where the objective is to understand the differences in network interactions.  This study has the potential to aid in the development of diagnostic markers and improve our understanding of such disease mechanisms.  The key contributions of the new methodology in terms of analyzing fMRI data include effective connectivity networks at each experimental time point (which to the best of our knowledge has not been possible before), and Granger causality networks inferred from the BTVT-VAR model.

The posterior distribution in large-dimensional tensor models is often multimodal, and standard MCMC algorithms may struggle to efficiently explore the entire posterior space, potentially getting stuck in local modes. Thus, we obtain posterior inference using an efficient Gibbs sampler with tempering, the Metropolis coupled Markov chain Monte Carlo ($\text{MC}^3$) algorithm proposed by \citet{altekar2004parallel}. By running parallel chains at different temperatures and allowing them to swap states, $\text{MC}^3$ allows for an improved exploration of the multimodal posterior distribution.\\
The remainder of the paper is organized as follows. In Section \ref{Model}, we formulate the time-varying tensor model and elucidate how to obtain dimension reduction via a tensor decomposition into a set of latent base matrices and binary indicators of connectivity patterns over time.    In Section \ref{IsingPrior}, we describe the Ising model specification on the temporal transitions, whereas in Section \ref{SparsityPriors} we describe the sparsity-inducing priors on the active elements of the tensor decomposition. In Section \ref{Posterior Computation} we discuss posterior computation and inference. Results of the simulation studies as well as the real data application are shown in Section \ref{Simulation Studies} and Section \ref{Real Data Application}, respectively.  Finally, Section \ref{Conclusion} provides some concluding remarks and future work.

\section{Time-Varying Tensor VAR model for Effective Connectivity}
\label{Model}

In this section, we introduce the proposed time-varying tensor VAR  specification for studying dynamic brain effective connectivity. Let $\mathbf{y}_t$ be an $N$-dimensional vector for $t=1,\dots,T$. Each time-series data $(y_{i1}, \ldots, y_{iT})$  represents the fMRI BOLD signal recorded at a voxel or region of interest (ROI) $i$, $i=1, \ldots, N$. 
The time-varying tensor VAR model of order $P$ assumes that $\mathbf{y}_t$ is a linear combination of the $P$ lagged signals  $\mathbf{y}_{t-1},\dots,\mathbf{y}_{t-P}$ plus an independent noise term, $\boldsymbol{\epsilon}_t\in \mathbb{R}^N$,
\begin{equation}
    \mathbf{y}_t=\begin{bmatrix}
    A_{1,t}, A_{2,t}, \dots, A_{P,t}    
    \end{bmatrix}
    \begin{bmatrix}
    \mathbf{y}_{t-1} \\
    \vdots \\
    \mathbf{y}_{t-P}
    \end{bmatrix}+\boldsymbol{\epsilon}_t,
\label{eq:TV-VAR}
\end{equation}
where $\boldsymbol{\epsilon}_t\sim \mathcal{N}(0,\Sigma)$ and the linear coefficients $A_{j,t}$, $j=1, \ldots, P$ are $N\times N$ matrices which are allowed 
to vary across $t$, $t=1,\dots,T$. We assume that $\Sigma$ is time-invariant and diagonal, and we focus on the coefficient matrices $\begin{bmatrix}A_{1,t}, A_{2,t}, \dots, A_{P,t} \end{bmatrix}$. If needed, the assumption on $\Sigma$ can be appropriately relaxed. The number of coefficients to be estimated  is $(T-P)\times N^2\times P+N$; hence, it is not possible to use the conventional ordinary least square estimator. We propose a model that induces a low-rank sparse representation of the coefficient matrices, implementing several concurrent strategies.

\subsection{Time-Varying Latent Component Representation of the VAR coefficient matrices.}

To start, we parsimoniously model the dynamic VAR coefficient matrices $[A_{1,t}, \ldots, A_{P,t}]$ as a \emph{time-varying composition} of $H$ latent static base matrices $[A^*_{1,h}, \ldots, A^*_{P,h}]$,
$h=1, \ldots, H$.  
Specifically,  for $h=1,\dots,H$, we introduce a binary-valued time series, $(\gamma_{h,t})_{t \geq P+1}$. Then, we assume
\begin{equation}
	\begin{bmatrix}
		A_{1,t}, A_{2,t}, \dots, A_{P,t}    
	\end{bmatrix}=\sum_{h=1}^H\gamma_{h,t}\begin{bmatrix}
		A_{1,h}^\ast, A_{2,h}^\ast, \dots, A_{P,h}^\ast
	\end{bmatrix},
\label{eq:VAR-reduction}
\end{equation}
that is,  at any given time $t$, the VAR coefficient matrices are built as a sum of a subset of base matrices $[A^*_{1,h}, \ldots, A^*_{P,h}]$, activated when $\gamma_{h,t}=1$. For instance, if at time $t$, $\gamma_{h,t}=1$ for $h=1, 2$ and zero for  $h=3, \ldots, H$, then $A_{j,t}= A^*_{j,1}+A^*_{j,2}$,  $j=1, \ldots, P$. However, our primary interest is not in the base matrices $[A^*_{1,h}, \ldots, A^*_{P,h}]$,  but rather in accurately recovering the time-varying VAR coefficients.  The binary $\gamma_{h,t}$'s  can be interpreted as indicators of latent individual or experimental conditions.  
Similarly, $A_{1,h}^\ast, A_{2,h}^\ast, \dots, A_{P,h}^\ast$ can be interpreted as {\it latent base  matrices}.

Hidden Markov Models (HMM) with time-invariant transition probabilities have been used in the neuroimaging literature to describe temporal variations in functional connectivity patterns \citep{Baker2014HMMfasttransient, Vidaurre12827, warnick2018bayesian}. Our proposed representation assumes that the dynamics of connectivity patterns result from the juxtaposition and cessation of the base connectivity matrices over time. An equivalent HMM would assume \(2^H\) states, leading to a \(2^H \times 2^H\) transition matrix, which would require estimating a prohibitively large number of parameters, even with moderate \(H\). The proposed approach leads to a more parsimonious representation and aligns more closely with current thinking among neuroscience investigators, as it reflects the belief that connectivity states do not undergo sudden changes over time. Instead, they emerge from the aggregation of multiple underlying states. Functional connectivity is thus understood as arising from latent variables derived from the shared variances in connectivity across task states \citep{Ou2014-bu, McCormick2022-qb}.

Compared to estimating \(N^2 \times P\) time-series of length \((T-P)\) in the initial model specification, our formulation \eqref{eq:VAR-reduction} requires estimating \(N^2 \times P\) base matrices. The dynamics of the VAR coefficient matrices are then governed by the temporal dependence between the \(\gamma_{h,t}\)'s, \(h=1, \ldots, H\), \(t=P+1, \ldots, T\). We assume independence among the \(\gamma_{h,t}\)'s across \(h=1, \ldots, H\), and model Markovian dependence across different time points through a latent Ising model that depends only on $2\times H$ parameters (see Section \ref{IsingPrior}).

\subsection{Rank-varying PARAFAC decomposition}
Despite the reduced dimensionality,  \eqref{eq:VAR-reduction} remains highly parameterized. Hence, we further propose to  stack each set of matrices $\begin{bmatrix}A_{1,h}^\ast, A_{2,h}^\ast, \dots, A_{P,h}^\ast\end{bmatrix}$ into a three-way tensor $\mathcal{A}_h^\ast$ of size $N\times N\times P$, and assume it can be decomposed as 
\begin{equation*}
\mathcal{A}_h^\ast=\alpha_{1,h}\circ\alpha_{2,h}\circ\alpha_{3,h},  \qquad h=1, \dots, H,
\end{equation*}
where $\circ$ indicates the vector outer product and $\alpha_{1,h}, \alpha_{2,h} \in \mathbb{R}^N$, $\alpha_{3,h}\in \mathbb{R}^P$ are the \emph{tensor margins} along the dimensions of the tensor. Thus, the time-varying VAR coefficients in \eqref{eq:VAR-reduction} can be stacked and be represented as the tensor
\begin{equation}
\mathcal{A}_t=\sum_{h=1}^H\, \gamma_{h,t}\,\mathcal{A}_h^\ast=\sum_{h=1}^H\gamma_{h,t}\,\left(\alpha_{1,h}\circ\alpha_{2,h}\circ\alpha_{3,h}\right).
\label{eq:rank-varying-PARAFAC}
\end{equation}
As the binary latent indicator $\gamma_{h,t}$ may be zero for some $t$, the decomposition \eqref{eq:rank-varying-PARAFAC} effectively defines a \emph{time-varying PARAFAC decomposition}, with an \emph{effective} rank that \emph{changes dynamically} over time and  is always less than or equal to $H$. Thus, at each time point, the time-varying tensor VAR coefficients are still represented as a combination of active subsets from the PARAFAC decomposition.

The PARAFAC decomposition, along with the Tucker decomposition, is often employed for tensor dimension reduction due to its straightforward interpretation and implementation. Tensors have  been utilized in the neuroimaging literature for detecting activations  via  tensor regression approaches \citep{zhou2013tensor, guhaniyogi2017bayesian}. However, to the best of our knowledge,  their use for studying dynamic effective connectivity  within VAR models has not yet been explored. For a brief introduction to the PARAFAC decomposition, we refer to Section S1 of the Supplementary Material \citep{zhang2024}.  

The number of parameters in \eqref{eq:rank-varying-PARAFAC} is $H(T-P)+H(2N+P)$, that is, linear in the observation size $N$, instead of $N^2$ without the tensor reparameterization. The decomposition is invariant under any permutation of the component indices $h$. In other words, for any permutation $\Pi(\cdot)$ of the index set $\{1,2, \ldots, H\}$,  $\mathcal{A}_t=\sum_{h=1}^H \, \gamma_{\Pi(h),t} \, \alpha_{1,\Pi(h)}\circ\alpha_{2,\Pi(h)}\circ\dots\circ\alpha_{M,\Pi(h)}$. Additionally, $\mathcal{A}_t$ is not affected by rescaling. Specifically, 
$\mathcal{A}_t=\sum_{=1}^R\alpha^*_{1,h}\circ\alpha^*_{2,h}\circ\dots\circ\alpha^*_{M,h}$, where  $\alpha^*_{m,r} = \nu_{m,r}\alpha_{m,r}$ for any set of multiplying factors $\nu_{m,r}$ such that $\prod_{m=1}^M \nu_{m,r} = 1$.  Indeed, it is important to note that while inference on the individual tensor margins may suffer from identifiability issues, the tensor $\mathcal{A}_t$ and the matrices $A_{j,t}$, $j=1,\ldots,P$ remain identifiable. Specifically, the temporal pattern of the VAR coefficients, obtained from \eqref{eq:VAR-reduction}--\eqref{eq:rank-varying-PARAFAC}, remains identifiable.

Using tensor algebra, we can rearrange the modes of the tensor decomposition in \eqref{eq:VAR-reduction} to express the base matrices as follows:
\begin{equation*}
\begin{bmatrix}A_{1,h}^\ast, A_{2,h}^\ast, \dots, A_{P,h}^\ast\end{bmatrix}=
\alpha_{3,h}' \otimes \left(\alpha_{1,h}\circ\alpha_{2,h}\right), \qquad h=1, \dots, H,
\end{equation*}
where $\otimes$ denotes the Kronecker product. The matricization operation highlights that the element-by-element ratio between $A_{1,h}^\ast$ and $A_{2,h}^\ast$ is proportional to the ratio between the first two entries of $\alpha_{3,h}$, and similarly for subsequent lags. After matricization, the set of time-varying  VAR coefficient matrices can be expressed as 
\begin{equation*}
\begin{bmatrix}
    A_{1,t}, A_{2,t}, \dots, A_{P,t}    
    \end{bmatrix}=\sum_{h=1}^H\, \gamma_{h,t}\, \alpha_{3,h}' \otimes \left(\alpha_{1,h}\circ\alpha_{2,h}\right),
\end{equation*}
where the latent binary time-series $\gamma_{h,t}$'s selects the active components at each time, $h=1, \ldots, H$.
The elements of the tensor margin $\alpha_{3,h} \in \mathbb{R}^P$, corresponding to the lag $P$ of the VAR model, play a crucial role in determining the weighting of the selected components. We expect the influence of past variables to diminish with increasing lags; therefore, it is reasonable to expect that the magnitude of the entries of $\alpha_{3,h}$ would also decrease as the lags increase. In Section \ref{SparsityPriors}, we describe prior specifications that promote sparsity in $\alpha_{1,h}$ and $\alpha_{2,h}$ while also enforcing increasing shrinkage in $\alpha_{3,h}$. 

\citet{sun2019dynamic} have also proposed the stacking of a series of dynamic tensors to form a higher order tensor. In their approach,  the data are observed tensors to be clustered over time  along the modes generated via the PARAFAC decomposition. Instead,  in our approach the data are multivariate time series and the tensor structure is used to construct a lower dimensional parameter space for the unknown VAR coefficients to be estimated.  Moreover, \citet{sun2019dynamic} achieve  smoothness in the parameters through a   fusion structure that penalizes discrepancies between neighboring entries in the same tensor margin. We follow a Bayesian approach and further encourage a contiguous structure by means of the Ising model specification detailed in the following section. 

\section{Latent Ising modeling for temporal transitions}
\label{IsingPrior}

The sequence of latent indicators $\gamma_{h,t}$ determines the time-varying activations of the %latent 
base matrices in the VAR model \eqref{eq:TV-VAR}--\eqref{eq:VAR-reduction}. To model this latent process, we propose an innovative use of the Ising model in the time domain. This approach encodes a Markovian dependence but allows modeling the time-varying activations as a function of only two parameters for each $h=1, \ldots, H$: one parameter captures general sparsity, and the other captures the strength of dependence between adjacent time points. 
  More specifically, we assume that, independently for each $h$, the binary state process $(\gamma_{h,t})_{t > P}$ is characterized by joint probability mass functions
\begin{align}
\begin{split}
		\pi\left(\gamma_{h,P+1},\dots,\gamma_{h,T}\right.&\left.\mid \theta_h,\kappa_h\right) \\ \propto&\exp\left(\theta_h\gamma_{h,P+1}+\sum_{t=P+2}^{T-1}\theta^*_h\gamma_{h,t}+\theta_h\gamma_{h,T}+\sum_{t=P+1}^{T-1}\kappa_h\gamma_{h,t}\gamma_{h,t+1}\right).
		\label{eq:Ising prior}
\end{split}
\end{align}
Equation \eqref{eq:Ising prior} defines an Ising model or an undirected graphical model or a Markov random field involving the binary random vector $\bm{\gamma}_h=(\gamma_{h,P+1}, \ldots, \gamma_{h,T}) \in\{0,1\}^{T-P}$, $h=1, \ldots, H$ \citep[see, e.g.,][]{WainwrightJordan2008}.
The parameters $\theta_h$ and $\theta_h^*$ can be interpreted as  {\em sparsity} parameters, since they correspond to the probability of activation for component $h$ at each time $t$, irrespective of the status at  $t-1$ and $t+1$. 
Positive values of $\theta_h$ and $\theta^*_h$ increase the probability that $\gamma_{h,t}=1$; on the other hand, negative values of $\theta_h$ and $\theta^*_h$  increase the probability that $\gamma_{h,t}=0$, $t=P+1, \ldots, T$. The parameter $\kappa_h$ captures the effect of the interaction between $\gamma_{h,t}$ and $\gamma_{h,t+1}$, that is, the persistence of activations over time. In particular, when $\kappa_h>0$, the probability that $\gamma_{h,t}$ and $\gamma_{h,t+1}$ are both non-zero is larger.

The Ising model \eqref{eq:Ising prior} can be seen as a specific instance of a multivariate Bernoulli distribution, as defined by \citet{dai2013multivariate}. In particular, in the following, we show that the proposed Ising process  is equivalent to a binary discrete autoregressive NDARMA$(1)$ model \citep{jacobs1983stationary, macdonald1997hidden, jentsch2019generalized}.
For notational simplicity, we focus on a single time series $\gamma_{h,t}$, and we omit the subscript $h$ for the remainder of the section. 
We start by recalling that a NDARMA(1) process is a binary time series that satisfies 
\begin{equation}
\label{eq:NDARMA}
    \gamma_t=a_t\, \gamma_{t-1}+(1-a_t) \,\epsilon_{t}, \quad t=1,\dots,T,
\end{equation}
where $a_t \stackrel{i.i.d.}{\sim} \text{Bern}(p_1)$, and $\epsilon_t \stackrel{i.i.d.}{\sim} \text{Bern}(p_2)$, with independent $a_t$ and $\epsilon_t$. The initial condition assumes $\gamma_1 \sim \text{Bern}(p_2)$.  
The NDARMA(1) model has a Markovian dependence structure, with transition probabilities
$$P(\gamma_t=1 \mid \gamma_{t-1})=p_1\, \mathbbm{1}(\gamma_{t-1}=1)+(1-p_1)\, p_2,$$
for $\gamma_t,\gamma_{t-1}\in\{0,1\}$. Moreover, marginally $\gamma_t \sim \text{Bern}(p_2)$. 
Intuitively, the autocorrelation function at lag 1 of the NDARMA time series is always positive, meaning that $\gamma_t$ and $\gamma_{t+1}$ tend to assume the same value, consistent with the contiguous behavior that the temporal-Ising model \eqref{eq:Ising prior} encourages when $\kappa$ is assumed to be positive.
Then, in a NDARMA$(1)$ model, the joint probability mass function of $\gamma_1,\dots,\gamma_T$ can be characterized a multivariate Bernoulli distribution \citep{dai2013multivariate}, as 
\begin{align*}
      P\,(\gamma_1,\dots,\gamma_T)&=p_{0\dots0}^{\prod_{t=1}^T(1-\gamma_t)}\, p_{10\dots0}^{\gamma_1\prod_{t=2}^T(1-\gamma_t)}
      \,p_{01\dots0}^{(1-\gamma_1)\,\gamma_2
      \,\prod_{t=3}^T(1-\gamma_t)}\cdots p_{1\dots1}^{\prod_{t=1}^T\gamma_t},
\end{align*}   
where $p_{0\dots0}$ indicates the probability of a zero sequence, that is $p_{0\dots0}=P(\gamma_1=0,\dots,\gamma_T=0)=(1-p_2)\,\prod_{t=2}^T\,(p_1 + (1-p_1)\,(1-p_2))$ and, similarly, for the other factors. Let $n=\sum_{t=1}^T \gamma_t$ indicate the total number of active indicators $\gamma_t$'s along the entire time-series. 
The following proposition maps the parameters $(\theta,\kappa)$ in the Ising model \eqref{eq:Ising prior} to the parameters $(p_1,p_2)$ in the NDARMA$(1)$ model in (\ref{eq:NDARMA}):
\begin{prop} \label{prop:Ising-NDARMA}
    The probability law of the NDARMA$(1)$ model in \eqref{eq:NDARMA} can be expressed as in \eqref{eq:Ising prior}. In particular, the parameters $(\theta,\kappa)$ are obtained as a function of the parameters $p_1$, $p_2$ in  \eqref{eq:NDARMA}  as 
    \begin{equation*}
        e^\theta=\frac{p_2(1-p_1)}{p_1+(1-p_2)(1-p_1)}, \quad e^\kappa=\frac{p_1+p_2(1-p_2)(1-p_1)^2}{p_2(1-p_2)(1-p_1)^2},
    \end{equation*}
    \begin{equation*}
        \exp(\theta^*)=\frac{p_2(1-p_2)(1-p_1)^2}{(p_1+(1-p_2)(1-p_1))^2}=\frac{e^\theta(e^\theta+1)}{e^{\theta+\kappa}+1}.
    \end{equation*}
    Inversely,
    \begin{equation*}
        p_1=\frac{e^\theta(e^\kappa-1)}{(e^{\theta+\kappa}+1)(e^\theta+1)}, \quad p_2=\frac{e^\theta(e^{\theta+\kappa}+1)}{e^{2\theta+\kappa}+2e^\theta+1}.
    \end{equation*}
    \label{prop1}
\end{prop}

\begin{figure}
    \centering
    \includegraphics[height=.35\linewidth]{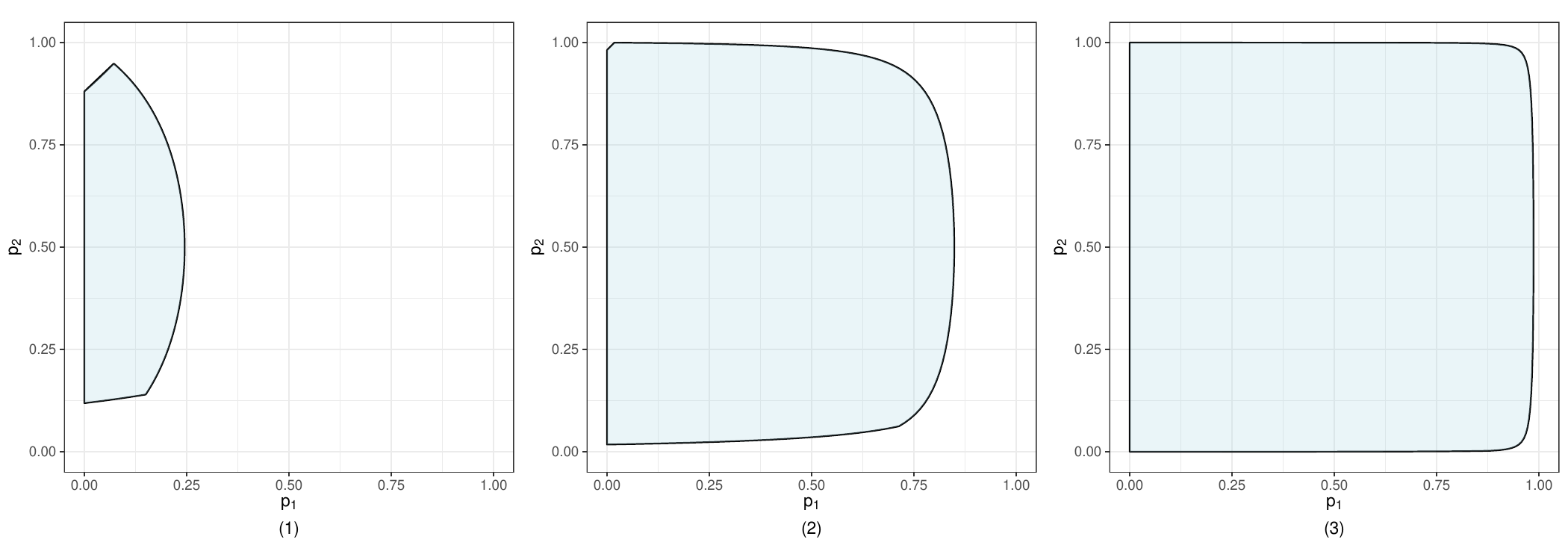}
    \caption{An illustration of the mapping between the parameters values  of $(p_1,p_2)$ of the NDARMA(1) model \eqref{eq:NDARMA} and the parameters $(\theta, \kappa)$ of the Ising model \eqref{IsingPrior}. The domain of $(\theta, \kappa)$ constrains the admissible range of $(p_1,p_2)$. The three panels illustrate the domain of $(p_1,p_2)$ corresponding to increasing domains of $(\theta, \kappa)$:  (1) $-2< \theta <2, 0<\kappa<1$; (2) $-4< \theta <4, 0<\kappa<5$; (3) $-8< \theta <8, 0<\kappa<10$.}
    \label{fig:Ising NDARMA}
\end{figure}
We refer to Section S2 of the supplementary material for the proof \citep{zhang2024}.

\subsection{Prior specification}

The result in Proposition \ref{prop:Ising-NDARMA} is helpful in setting the prior distributions for the parameters in \eqref{eq:Ising prior}, as it highlights underlying constraints among the parameters and how the domain of $(\theta, \kappa)$ constrains the admissible range of $(p_1,p_2)$. Indeed, the transformation is bijective, since $\theta^*$ can be expressed as a function of the pair $(\theta, \kappa)$.  The parameters $\kappa$ and $p_1$ have the same sign, since they both indicate the strength of the dependence between two neighboring $\gamma_{t-1}$ and $\gamma_t$. It also follows that, given that   $\kappa$ is positive, $\theta^*$ is always smaller than $\theta$. As an illustration of the complex dependencies induced by the mapping between $(\theta, \theta^*, \kappa)$ in \eqref{eq:Ising prior} and $(p_1, p_2)$ in \eqref{eq:NDARMA},  Fig \ref{fig:Ising NDARMA} compares the range of $(p_1,p_2)$ for different intervals of values of $(\theta,\kappa)$. For example, if $-2<\theta<2$ and $0<\kappa<1$, the corresponding set of NDARMA(1) models is limited to a subset of those with $p_1<0.25$, whereas if $-8<\theta<8$ and $0<\kappa<10$, the support covers almost all values of $p_1$ and $p_2$ between 0 and 1. As the domain of $(\theta,\kappa)$ expands, the set of induced NDARMA(1) models also expands. Shrinking $\gamma_{t}$ towards zero is desirable for regularization purposes, which corresponds to allowing negative values of the parameter $\theta$. 
However, excessive shrinkage may hinder our ability to identify recurrent latent base patterns, as it can lead to  very low estimates of $p_1$ and $p_2$. Indeed, it is well known that the prior specification of the parameters of a Ising model needs to be  conducted with care, in order to avoid the phenomenon of phase-transition \citep{Li2010, Li2015}. In statistical physics,  a phase-transition refers to a sudden change from a disordered (non-magnetic) to an ordered (magnetic) state at low temperatures. In Bayesian variable selection, the phase-transition has been associated to values of the parameter space that lead to selecting either all or none of the tested variables. These considerations motivate our suggestion of a proper uniform prior distribution on the parameters $(\theta,\kappa)$ over a closed interval in $\mathbb{R}^2$. More specifically, we assume that, for each $h=1, \ldots, H$,   $\theta_h$ lies between $\left[\theta_{h,\min},\theta_{h,\max}\right]$ with lower limit $\theta_{h,\min} <0$ and upper limit $\theta_{h,\max} >0$. We have found that choosing $\theta_{h,\min}=-8$ and $\theta_{h,\max}=8$ ensures a proper exploration of the parameter space and appears to avoid phase transitions. Similarly, for $\kappa_h$, we encourage a contiguous structure where $\gamma_{h,t}$ and $\gamma_{h,t+1}$ are simultaneously selected by assuming that $\kappa_h$ is positive with a uniform prior on   $\kappa_h\in\left[0,\kappa_{h,\max}\right]$ with $\kappa_{h,\max}>0$. Also, in this case,  an upper limit $\kappa_{h,\max}=10$ appears to ensure both reasonably good
inference on the  time-varying coefficients and computational efficiency. 

\section{Sparsity-inducing priors on the tensor components}
\label{SparsityPriors}

In addition to achieving dimension reduction through the PARAFAC decomposition, we further seek  shrinkage of the tensor margins' parameters. Hence, we consider priors that shrink the parameters toward zero, enabling a sparse representation of the VAR coefficients and more interpretable estimation of the connectivity patterns. In particular, for the elements of the tensor margins $\alpha_{1,h}$ and $\alpha_{2,h}$ we consider a multi-way Dirichlet generalized  double Pareto prior \citep{guhaniyogi2017bayesian}, whereas for the lag margin $\alpha_{3,h}$, we consider an increasing shrinkage prior, so that higher-order lags are penalized. 

More specifically, we assume that the $N$-dimensional tensor margins $\alpha_{1,h}$ and $\alpha_{2,h}$ are  normally distributed with zero mean and variance-covariance matrix  $\tau\phi_h W_{l,h}$, with  $W_{l,h}=\text{diag}(\{w_{l,h,k}\}_{k=1}^N)$, $l=1,2$. The parameter $\tau$ is a global scale parameter that follows a  Gamma distribution, whereas $\phi_1,\dots,\phi_H$ are local scale parameters that follow a symmetric Dirichlet distribution. The diagonal elements of the covariance matrix have a generalized double Pareto prior. Thus,
\begin{align*}
	\alpha_{l,h} \mid \phi_h, \tau, W_{l,h} &\sim \mathcal{N}(0,\phi_h\tau W_{l,h}), \quad l=1,2\\
 w_{l,h,k} \mid \lambda_{l,h} &\sim \text{Exp}\left({\lambda_{l,h}^2}/2\right) \quad 1\leq k \leq N, l=1,2\\
\lambda_{1,h}, \lambda_{2,h} & \stackrel{i.i.d.}{\sim} \text{Ga}(a_\lambda,b_\lambda) \\ 
	\phi_1,\dots,\phi_H &\sim \text{Dirichlet}(a, \dots, a),\\ \tau &\sim \text{Ga}(a_\tau,b_\tau).
\end{align*} 
By setting $a_\tau=H a$, one can obtain tractable full conditionals distributions  for $\tau$ and $(\phi_1,\dots,\phi_H)$, since the full conditional for $\tau$ is a generalized inverse Gaussian distribution and the full conditionals for $(\phi_1,\dots,\phi_H)$ are normalized  generalized inverse Gaussian random variables \citep{guhaniyogi2017bayesian}. 

Also for the lag mode parameter, $\alpha_{3,h}$, we employ a normal prior with a global-local structure on the covariance, specifically, $\alpha_{3,h} \mid \phi_h, \tau, W_{3,h} \sim \mathcal{N}(0,\phi_h\tau W_{3,h})$. However, we incorporate a cumulative shrinkage effect on the diagonal entries of $w_{3,h}$ to encourage the estimation of a small number of lags. More specifically, we use a  cumulative shrinkage prior \citep{legramanti2020bayesian}. This type of prior induces increasing shrinkage through a series of spike and slab distributions, which assign greater mass to a target spike value as model complexity increases. Let $W_{\infty}$ indicate the target spike. Here, we set $W_{\infty}=0.01$  to prevent the Normal distribution from degenerating into a point mass and improve computational efficiency \citep{GeorgeMcCulloch1993}. Then, for $1\leq j \leq P$, we assume that the diagonal elements of $W_{3,h}$,
\begin{align*}
    w_{3,h,j} \mid z_{h,j} \sim [1-\mathbbm{1}(z_{h,j}\, \leq j)]\, \text{InvGa}(a_w,b_w) + \mathbbm{1}(z_{h,j}\leq j)\, \delta_{W_\infty},
    \end{align*}
where each $z_{h,j}$ is a draw from a Multinomial  random variable with $\text{pr}(z_{h,j}=l\mid \omega_{h,l})=\omega_{h,l}$, for $l=1, \ldots, P$.
Thus, the probability of selecting the target spike is increasing with the lags $j$, since $P(z_{h,j}\leq j)=\sum_{l=1}^j \omega_{h,l}$. The weights $\omega_{h,l}$ are obtained through a stick-breaking construction \citep{Sethu94}, that is, $\omega_{h,l} = v_{h,l}\prod_{m=1}^{j-1}(1-v_{h,m})$, $v_{h,l} \sim \text{Beta}(\beta_1,\beta_2)$, $1\leq l \leq P$.  Correspondingly, the probability of choosing the  Inverse Gamma slab component is $P(z_{h,j}> j)=\prod_{l=1}^j (1-v_{h,l})$, which decreases with $j$. Higher sparsity levels for the modes $\alpha_{1,h}, \alpha_{2,h}$ and $\alpha_{3,h}$ are obtained by setting smaller values of $a_\tau$ relative to $b_\tau$ or by setting smaller values of $b_\lambda$ relative to $a_\lambda$. This is evident from the expectations $\text{E}(\tau)=a_\tau/b_\tau$ and $\text{E}(W_{1,h,k})=2b_\lambda^2/[(a_\lambda-1)(a_\lambda-2)]$.
We discuss these choices in Section \ref{Posterior Computation}.

\subsection{Rank of the PARAFAC decomposition}

A crucial point in the representation \eqref{eq:rank-varying-PARAFAC} is the choice of the rank, $H$. One widely adopted option is to regard this choice as a model selection problem and naturally resort to information criteria such as AIC or BIC \citep{zhou2013tensor, wang2021high, wang2016modeling,davis2016sparse}. \citet{luo2022bayesian} use an increasing shrinkage prior for inference on the rank. Here, we propose a different strategy. Drawing from recent Bayesian literature on overfitted mixture models  \citep[see, e.g.,][]{MalsinerWalli2016, RousseauMengersen2011}, we set the parameters of the  sparsity-inducing hierarchical prior in Section \ref{SparsityPriors} to automatically shrink unnecessary components to zero.  More specifically,  a small concentration parameter $a$ of  the symmetric Dirichlet distribution  $(\phi_1,\dots,\phi_H)$ assigns more probability mass to the edges of the simplex, meaning that more components become redundant.  As a result, within the available $H$ components, only a few will be effectively different from zero. Additionally, we can utilize the cumulative shrinkage prior for the VAR lag order $P$ for further regularization,  by selecting appropriate values for $\beta_1, \beta_2, a_w$ and $b_w$. For instance, a large value of $\beta_1$ and a small $\beta_2$ encourage a more parsimonious VAR model by assigning little probability to higher orders. Therefore, by accepting a modest increase in computational demand,  %at the expense of a slightly higher computational demand, 
it is possible to  set relatively high values for $H$ (and $P$),  and then rely on the implicit regularization of the shrinkage priors to determine the actual number of effective components (and lags), without the need for ranking different models in practice. Fig \ref{fig:hierarchical diagram} summarizes  the proposed hierarchical model on the $N$-dimensional time series $\bm y_t$ using a directed graph representation.

\begin{figure}[!ht]
\centering
\begin{tikzpicture}
\tikzset{box/.style = {rectangle, inner sep=0pt, text width=10mm, text height=-5mm, font=\tiny}}
\tikzset{vertex1/.style = {shape=circle,draw,minimum size=1.5em}}
\tikzset{vertex2/.style = {shape=rectangle,draw,minimum size=1.5em}}
\tikzset{vertex3/.style = {shape=rectangle,draw,dashed,minimum size=1.5em}}
\tikzset{vertex4/.style =
{shape=circle,draw,fill=gray!50,minimum size=1.5em}}
\tikzset{edge/.style = {->,> = latex}}
\tikzset{fit margins/.style={/tikz/afit/.cd,#1,/tikz/.cd,inner xsep=\pgfkeysvalueof{/tikz/afit/left}+\pgfkeysvalueof{/tikz/afit/right},inner ysep=\pgfkeysvalueof{/tikz/afit/top}+\pgfkeysvalueof{/tikz/afit/bottom},xshift=-\pgfkeysvalueof{/tikz/afit/left}+\pgfkeysvalueof{/tikz/afit/right},yshift=-\pgfkeysvalueof{/tikz/afit/bottom}+\pgfkeysvalueof{/tikz/afit/top}},afit/.cd,left/.initial=2pt,right/.initial=2pt,bottom/.initial=2pt,top/.initial=2pt}

\node[vertex2] (cons1) at (0.46,0) {$a_\lambda$};
\node[vertex2] (cons2) at (1.04,0) {$b_\lambda$};
\node[vertex2] (cons3) at (3,0) {$a_w$};
\node[vertex2] (cons4) at (3.62,0) {$b_w$};
\node[vertex2,font=\scriptsize] (cons5) at (4.5,0) {$\!W_{\!\infty\!}$};
\node[vertex2] (cons6) at (7.51,-1.5) {$\beta_1$};
\node[vertex2] (cons7) at (8.09,-1.5) {$\beta_2$};

\node[vertex1,font=\scriptsize] (plate1_in1) at (0,-1.5) {$\!\lambda_{\!1\!,\!h\!}$};
\node[vertex1,font=\scriptsize] (plate1_in2) at (1.5,-1.5) {$\!\lambda_{\!2\!,\!h\!}$};
\node[vertex1] (plate1_in3) at (4,-1.5) {$\!z_{\!h\!,\!k}\!$};
\node[vertex1] (plate1_in4) at (6,-1.5) {$\!v_{\!h\!,\!k}\!$};
\node[vertex1,font=\scriptsize] (plate1_in5) at (0,-3) {$\!W_{\!1\!,\!h\!,\!k\!}$};
\node[vertex1,font=\scriptsize] (plate1_in6) at (1.5,-3) {$\!W_{\!2\!,\!h\!,\!k\!}$};
\node[vertex1,font=\scriptsize] (plate1_in7) at (4,-3) {$\!W_{\!3\!,\!h\!,\!k\!}$};
\node[vertex1,font=\scriptsize] (plate1_in8) at (0,-4.5) {$\!\alpha_{\!1\!,\!h\!,\!k\!}$};
\node[vertex1,font=\scriptsize] (plate1_in9) at (1.5,-4.5) {$\!\alpha_{\!2\!,\!h\!,\!k\!}$};
\node[vertex1,font=\scriptsize] (plate1_in10) at (4,-4.5) {$\!\alpha_{\!3\!,\!h\!,\!k\!}$};
\node[box] (plate1_text) at (6,-5) {$\!h\!=\!1\!,\!\sdots\!,\!H,$\\ $\!k\!=\!1\!,\!\sdots\!,\!N$};
\node[vertex3,fit margins={left=-6pt,right=5.5pt,bottom=2pt,top=0pt},fit=(plate1_in1) (plate1_in2) (plate1_in3) (plate1_in4) (plate1_in5) (plate1_in6) (plate1_in7) (plate1_in8) (plate1_in9) (plate1_in10) (plate1_text)] (plate1) at (2.6,-3) {};

\node[vertex2] (cons8) at (7.51,-3) {$a_\tau$};
\node[vertex2] (cons9) at (8.09,-3) {$b_\tau$};
\node[vertex2] (cons10) at (9,-3) {$a$};

\node[vertex1] (rd1) at (7.8,-4.5) {$\!\tau\!$};

\node[vertex1] (plate2_in) at (9,-4.5) {$\!\phi_h\!$};
\node[box] (plate2_text) at (9.7,-5) {$\!h\!=\!1\!,\!\sdots\!,\!H$};
\node[vertex3,fit margins={left=-7pt,right=5.5pt,bottom=1pt,top=0pt},fit=(plate2_in) (plate2_text)] (plate2) at (8.9,-4.5) {};

\node[vertex2] (cons11) at (0.52,-6) {$\!b_{\!\sigma\!}$};

\node[vertex1] (plate3_in) at (1.7,-6) {$\!\sigma^{\!2}_{\!k\!}$};
\node[box, below right=10mm of plate3_in.north west] (plate3_text) { $\!k\!=\!1\!,\!\sdots\!,\!N$};
\node[vertex3,fit margins={left=-6pt,right=5.5pt,bottom=0.5pt,top=0.5pt},fit=(plate3_in) (plate3_text)] (plate3) at (1.7,-6) {};

\node[vertex4] (obs) at (5,-6) {$\mathbf{y}_t$};

\node[vertex1] (plate45_in) at (6.5,-6) {$\!\gamma_{h\!,\!t\!}$};
\node[box, below right=8mm of plate45_in.west] (plate4_text) { $\!t\!=\!1\!,\!\sdots\!,\!T$};
\node[vertex3,fit margins={left=-6pt,right=4.5pt,bottom=2.5pt,top=2pt},fit=(obs) (plate45_in) (plate4_text)] (plate4) at (5.7,-6) {};

\node[vertex1] (plate5_in1) at (8,-6) {$\!\theta_{\!h\!}$};
\node[vertex1] (plate5_in2) at (6.5,-7.1) {$\!\kappa_{\!h\!}$};
\node[box] (plate5_text) at (8,-7.5) {$\!h\!=\!1\!,\!\sdots\!,\!H$};
\node[vertex3,fit margins={left=0pt,right=0pt,bottom=0.8pt,top=0.5pt},fit=(plate45_in) (plate5_in1) (plate5_in2) (plate5_text)] (plate5) at (7.25,-6.55) {};

\node[vertex2] (cons12) at (9.1,-6) {$\!\theta_{\!\min\!}$};
\node[vertex2] (cons13) at (9.9,-6) {$\!\theta_{\!\max\!}$};
\node[vertex2] (cons14) at (9.2,-7.1) {$\!\kappa_{\!\min\!}$};
\node[vertex2] (cons15) at (10.01,-7.1) {$\!\kappa_{\!\max\!}$};

\draw[edge] (0.75,-0.26) to (plate1_in1);
\draw[edge] (0.75,-0.26) to (plate1_in2);
\draw[edge] (plate1_in1) to (plate1_in5);
\draw[edge] (plate1_in2) to (plate1_in6);
\draw[edge] (plate1_in5) to (plate1_in8);
\draw[edge] (plate1_in6) to (plate1_in9);
\draw[edge] (3.32,-0.26) to (plate1_in3);
\draw[edge] (cons5) to (plate1_in3);
\draw[edge] (plate1_in4) to (plate1_in3);
\draw[edge] (plate1_in3) to (plate1_in7);
\draw[edge] (plate1_in7) to (plate1_in10);
\draw[edge] (cons6) to (plate1_in4);
\draw[edge] (7.8,-3.26) to (rd1);
\draw[edge] (cons10) to (plate2_in);
\draw[edge] (plate1_in8) to (obs);
\draw[edge] (plate1_in9) to (obs);
\draw[edge] (plate1_in10) to (obs);
\draw[edge] (rd1) to (obs);
\draw[edge] (plate2_in) to (obs);
\draw[edge] (cons11) to (plate3_in);
\draw[edge] (plate3_in) to (obs);
\draw[edge] (plate45_in) to (obs);
\draw[edge] (plate5_in1) to (plate45_in);
\draw[edge] (plate5_in2) to (plate45_in);
\draw[edge] (cons12) to (plate5_in1);
\draw[edge] (cons14) to (plate5_in2);

\end{tikzpicture}
\caption{A schematic directed graph representation of the hierarchical Bayesian time-varying tensor VAR model. In particular, the graph summarizes the Ising model on the latent selection indicators $\gamma_{h,t}$ and the regularization priors on the tensor margins $\alpha_{1,h}$, $\alpha_{2,h}$, $\alpha_{3,h}$ of the PARAFAC decomposition of the base matrices $A^*_{h,j}$, $h=1,\ldots, H$, $j=1, \ldots, P$.} \label{fig:hierarchical diagram}
\end{figure}

\section{Posterior Computation}
\label{Posterior Computation} 

To infer the dynamic coefficient matrices and latent indicators in our time-varying VAR model, we utilize Markov Chain Monte Carlo (MCMC) methods. Despite the initially daunting number of parameters, our approach reduces complexity through tensorization, dynamic composition of latent base matrices, and the temporal Ising process. However, an accurate use of MCMC methods is essential due to the complexity, multidimensional and dynamic nature of the time series.

The prior specification allows the use of a blocked Gibbs sampler to draw samples from the posterior distribution. When sampling from the posterior distribution for $\theta_h$ and $\kappa_h$ using a Metropolis-Hastings algorithm, the normalizing constant depends on the sampled parameters, giving rise to a well-known issue of sampling from a doubly-intractable distribution. Thus, we follow the auxiliary variable approach proposed by \cite{moller2006efficient} to obtain the posterior samples from the Ising model. More specifically, the approach introduces an auxiliary variable such that -- by  adding its full conditional to the Metropolis-Hasting ratio --  the normalizing constant is canceled out. 
Posterior samples of the latent auxiliary variable are then obtained using the equivalent NDARMA representation which, remarkably, improves efficiency. The use of the auxiliary variable approach is also key for allowing the update of the $\gamma_{h,t}$'s for all $t=P+1,\dots,T$ and each $h=1,\ldots, H$. 

In our fMRI data application, to fully explore the geometry of the target posterior distribution, we implement the Metropolis coupled MCMC algorithm on top of the blocked Gibbs sampler. The core idea is to raise the BTVT-VAR likelihood to a sequence of descending powers $\rho$, resulting in multiple different posteriors. When $\rho=1$, it is the target distribution and when $\rho=0.1$, it corresponds to the almost flat posterior. The flatter the power posterior is, the easier it is to sample the whole surface. The multiple MCMCs, each assigned with a specific power $\rho$, are then run in parallel. At the end of every pre-fixed iteration, we preform a swap step where we randomly choose one chain with $\rho<1$, calculate the Metropolis-Hastings ratio between the chosen chain and the target chain based on the last draws and decide whether or not to exchange them using the Metropolis-Hastings updating rule. The details of the Gibbs sampler and the $\text{MC}^3$ algorithm are reported in S3 and S4 of the supplementary material \citep{zhang2024}.

Through MCMC samples from the posterior distribution we can obtain inferences on the dynamic coefficient matrices $\begin{bmatrix}A_{1,t}, A_{2,t}, \dots, A_{P,t}\end{bmatrix}$. 
Posterior expectations at each time point are approximated by the MCMC averages.
We summarize inference on the  $\gamma_{h,t}$'s by computing the posterior mode, specifically we set $\Tilde{\gamma}_{h,t}=1$ whenever the posterior probability of activation, $P(\gamma_{h,t}=1\mid y_1, \ldots, y_T)$, 
is over 0.5.

\section{Simulation Studies}
\label{Simulation Studies}

We illustrate the performances of the proposed model formulation in two simulation studies.  

\emph{Simulation 1: Synthetic data}. 
We generate multivariate time series from a TVT-VAR model where the coefficients are estimated from the first two sessions of a subject's fMRI data during the reading experiment detailed in Section \ref{Real Data Application}. Specifically, we obtain the OLS estimates from a static VAR model with $N=27$ and $P=3$, fitting the two sessions separately. For each estimated coefficient matrix, we stack its elements and apply a rank-2 PARAFAC decomposition. The PARAFAC decomposition yields  $H=4$ different sets of tensor margins $\alpha_{1,h}, \alpha_{2,h}, \alpha_{3,h}$, which are then used to construct the time-varying coefficients. We generate 30 replicates for each dataset, with specific tensor margins sampled from normal distributions centered on $\alpha_{1,h}, \alpha_{2,h}, \alpha_{3,h}$, obtained from the real fMRI data as outlined above. The sampled tensor margins are further thresholded at 0.3 to induce a desirable level of sparsity. For each of the 30 replicates, we then generate a multivariate time course of length $T=400$ from a TV-VAR model with coefficients changing at time points 80, 160, 240, and 320, and assuming $H=4$, $P=3$. Within each time interval, the coefficients are assumed to be the sum of a time-invariant subset of the four base matrices calculated from the tensor margins. 

The error term $\boldsymbol{\epsilon}_t$ entries have identical standard deviations determined by the estimated signal-to-noise ratio (SNR) in the real data. Several SNR definitions exist in the literature \citep{welvaert2013definition, zhang2014spatio, haslbeck2021tutorial}. Following \citet{haslbeck2021tutorial}, we divide the estimated VAR coefficients' maximum size by the estimated standard deviation of the error term, obtaining SNR values of 1.8518 and 2.7972 for the two runs. In the data generation, we set the SNR to 2.5. To assess the robustness of the proposed Bayesian model, we also consider two additional scenarios with SNR values of 0.5 and 10. Detailed results for these cases are provided in Section S5 of the supplementary material \citep{zhang2024}. For model fitting, we set the hyper-parameter values to be $a_\lambda=3,b_\lambda=1,a=1/H,a_\tau=Ha=1,b_\tau=H^4,\beta_1=1,\beta_2=5,a_w=b_w=2,W_\infty=0.01,a_\sigma=10,b_\sigma=2,\theta_{h,\min}=-8, \theta_{h,\max}=8,\kappa_{h,\max}=10$, to encourage sparsity. A total of 10,000 MCMC samples are drawn, one-third of the output is discarded, and the remaining samples are thinned by a factor of 3 to reduce storage and possible auto-correlation of the chains. 

First, we compare the proposed method's performance under different combinations of $P$ and $H$. We summarize the results of the simulation study by investigating the ability of our model to recover the dynamic coefficient matrices $\begin{bmatrix}A_{1,t}, A_{2,t}, \dots, A_{P,t}\end{bmatrix}$ in \eqref{eq:VAR-reduction}. To assess the performance of the proposed method, we employ the root mean square Frobenius distance between the MCMC estimates of the posterior means \(E\left(A_{j, t} \mid y_{1}, \ldots, y_{T}\right)\) at each $t$, say $\{\Tilde{A}_{j,t}\}_{j=1,\dots,P, t=P+1,\dots,T}$, and the true matrices defined by 
\begin{equation}
    \text{err}(\{\Tilde{A}_{j,t}\}) = \sqrt{\frac{\sum_{t=P+1}^T\sum_{j=1}^P \norm{\Tilde{A}_{j,t}-A_{j,t}}^2_F }{(T-P)\times N^2P}}.
    \label{eq:performance_At}
\end{equation}
We also compare the obtained Bayesian point estimates with those provided by a frequentist time-varying vector auto-regressive model, as implemented in the R package \textbf{TvReg} \citep{tvReg1} and a mixed VAR model in the R package \textbf{mgm} \citep{haslbeck2020mgm}. The results for the SNR$=2.5$ are shown in Table \mbox{\ref{tab:components-performance-2}}. When $P$ and $H$ increase, performance usually improves. But if $P$ and $H$ are higher than the true values, performance doesn't improve significantly. For instance, the estimation errors of all entries in the time-varying coefficient matrices $A_{j,t}$ stay between 0.0263 and 0.0317 for different combinations of $P\geq 3, H\geq 4$. 
This observation showcases the effectiveness of the sparsity-inducing priors applied to the tensor margins $\alpha_{1,h}, \alpha_{2,h}$, and $\alpha_{3,h}$. 
The BTVT-VAR model appears to provide an improved estimation of the true dynamic structure of the data with respect to the non-sparse frequentist VAR.  In comparison to mixed VAR estimates obtained through elastic-net regularized neighborhood regression,  our method performs better in terms of recovering non-zero entries while maintaining sufficient sparsity over true zero entries. Table \ref{tab:components-performance-2} also shows the point estimates of 
the base matrices $[A_{1,h}^\ast, A_{2,h}^\ast, \dots, A_{P,h}^\ast]$. The error of the MCMC-based estimates of the  posterior means say  $\{\Tilde{A}_{j,h}^\ast\}_{j=1,\dots,P}$, is  similarly defined as
\begin{equation}
    \text{err}(\{\Tilde{A}_{j,h}^\ast\})=\sqrt{\frac{\sum_{j=1}^P\norm{\Tilde{A}_{j,h}^\ast-A_{j,h}^\ast}^2_F}{N^2P}}.
    \label{eq:performance_components}
\end{equation}
In order to compute \eqref{eq:performance_components}, we first need to address label switching \citep{Stephens2000}, which may affect the estimation of the components in \eqref{eq:VAR-reduction}. Thus, we develop a procedure to match the posterior means of each component to their true counterparts. Additionally, if the assumed value of $H$ exceeds the true value, some posterior mean estimates ${\Tilde{A}{j,h}^\ast}{j=1,\dots,P}$ might contain redundant elements very close to zero. To identify these essentially ``empty'' components, we consider the maximum norm $\max_{j=1,\dots,P} , \norm{\Tilde{A}{j,h}^\ast}_\infty$. If this norm is below a specified threshold, we set the component to $\bm 0$ and exclude it from further analysis. In the following, we set such threshold to 0.01. Next, to match the remaining posterior mean estimates $\{\Tilde{A}_{j,h}^\ast\}_{j=1,\dots,P}$ with the true components, we rank them based on ascending values of the Frobenius distances. The simulation results of the components indicate that it is sufficient to set $P$ and $H$ to sufficiently large values and let the shrinkage priors automatically select the most suitable $P$ and $H$.

For the inference on the latent binary indicators \(\left(\gamma_{h , P+1}, \ldots, \gamma_{h , T}\right), h=1, \ldots, H\), we threshold the estimated posterior probability of activation at each time point for each  data set to identify the activated $\gamma_{h,t}$'s  from the MCMC samples, as described in Section \ref{Posterior Computation}. Table \mbox{\ref{tab:gamma-performance-2}} shows  the average accuracy, sensitivity, specificity and precision across all 30 data sets. The results show that the model is able to reconstruct the components  and their dynamic activation reasonably well.

\captionsetup{width=\textwidth}
\begin{table}
\caption{\label{tab:components-performance-2} Simulation study 1; SNR = 2.5. Mean and standard deviation (in parentheses) of estimation errors over 30 iterations for posterior means of tensor components [$A_{j,h}^*$] and dynamic coefficients [$A_{j,t}$] in the BTVT-VAR model. Compared with frequentist estimates from the \textbf{tvReg} R-package \citep{tvReg1} and mixed models from the \textbf{mgm} R-package \citep{haslbeck2020mgm}. See Section 6 for details. \\}

\centering
\resizebox{\textwidth}{0.4\textheight}{
\begin{tabular}{|c|cccccc|}
\hline
\multirow{3}{*}{SNR=2.5} & \multicolumn{6}{c|}{BTVT-VAR} \\
\cline{2-7}
& \multicolumn{3}{c|}{Latent {$A_{j,h}^\ast$}} & \multicolumn{3}{c|}{VAR Matrices {$A_{j,t}$}} \\
\cline{2-7}
& All entries & True non-zeroes   & \multicolumn{1}{l|}{True zeroes } & All entries & True non-zeroes  & \multicolumn{1}{l|}{True zeroes} \\
\hline
P=2 H=3 &
\begin{tabular}[c]{@{}c@{}}0.0405\\ (0.0308)\end{tabular} & \begin{tabular}[c]{@{}c@{}}0.1977\\ (0.1641)\end{tabular} & \multicolumn{1}{c|}{\begin{tabular}[c]{@{}c@{}}0.0170\\ (0.0314)\end{tabular}} &
\begin{tabular}[c]{@{}c@{}}0.0541\\ (0.0206)\end{tabular} & \begin{tabular}[c]{@{}c@{}}0.1971\\ (0.0813)\end{tabular} & \begin{tabular}[c]{@{}c@{}}0.0216\\ (0.0153)\end{tabular} \\
\hline
P=2 H=4 &
\begin{tabular}[c]{@{}c@{}}0.0348\\ (0.0360)\end{tabular} & \begin{tabular}[c]{@{}c@{}}0.1567\\ (0.1366)\end{tabular} & \multicolumn{1}{c|}{\begin{tabular}[c]{@{}c@{}}0.0148\\ (0.0263)\end{tabular}} &
\begin{tabular}[c]{@{}c@{}}0.0475\\ (0.0165)\end{tabular} & \begin{tabular}[c]{@{}c@{}}0.1715\\ (0.0616)\end{tabular} & \begin{tabular}[c]{@{}c@{}}0.0187\\ (0.0071)\end{tabular} \\
\hline
P=2 H=5 &
\begin{tabular}[c]{@{}c@{}}0.0373\\ (0.0384)\end{tabular} & \begin{tabular}[c]{@{}c@{}}0.1651\\ (0.1519)\end{tabular} & \multicolumn{1}{c|}{\begin{tabular}[c]{@{}c@{}}0.0160\\ (0.0277)\end{tabular}} &
\begin{tabular}[c]{@{}c@{}}0.0485\\ (0.0226)\end{tabular} & \begin{tabular}[c]{@{}c@{}}0.1737\\ (0.0832)\end{tabular} & \begin{tabular}[c]{@{}c@{}}0.0209\\ (0.0123)\end{tabular} \\
\hline
P=3 H=3 &
\begin{tabular}[c]{@{}c@{}}0.0317\\ (0.0450)\end{tabular} & \begin{tabular}[c]{@{}c@{}}0.1522\\ (0.1796)\end{tabular} & \multicolumn{1}{c|}{\begin{tabular}[c]{@{}c@{}}0.0137\\ (0.0380)\end{tabular}} &
\begin{tabular}[c]{@{}c@{}}0.0450\\ (0.0231)\end{tabular} & \begin{tabular}[c]{@{}c@{}}0.1647\\ (0.1039)\end{tabular} & \begin{tabular}[c]{@{}c@{}}0.0194\\ (0.0126)\end{tabular} \\
\hline
P=3 H=4 &
\begin{tabular}[c]{@{}c@{}}0.0190\\ (0.0300)\end{tabular} & \begin{tabular}[c]{@{}c@{}}0.0856\\ (0.1384)\end{tabular} & \multicolumn{1}{c|}{\begin{tabular}[c]{@{}c@{}}0.0091\\ (0.0169)\end{tabular}} &
\begin{tabular}[c]{@{}c@{}}0.0317\\ (0.0233)\end{tabular} & \begin{tabular}[c]{@{}c@{}}0.1027\\ (0.0846)\end{tabular} & \begin{tabular}[c]{@{}c@{}}0.0173\\ (0.0095)\end{tabular} \\
\hline
P=3 H=5 &
\begin{tabular}[c]{@{}c@{}}0.0149\\ (0.0237)\end{tabular} & \begin{tabular}[c]{@{}c@{}}0.0622\\ (0.1013)\end{tabular} & \multicolumn{1}{c|}{\begin{tabular}[c]{@{}c@{}}0.0084\\ (0.0127)\end{tabular}} &
\begin{tabular}[c]{@{}c@{}}0.0275\\ (0.0176)\end{tabular} & \begin{tabular}[c]{@{}c@{}}0.0875\\ (0.0728)\end{tabular} & \begin{tabular}[c]{@{}c@{}}0.0161\\ (0.0054)\end{tabular} \\
\hline
P=4 H=3 &
\begin{tabular}[c]{@{}c@{}}0.0295\\ (0.0348)\end{tabular} & \begin{tabular}[c]{@{}c@{}}0.1496\\ (0.1765)\end{tabular} & \multicolumn{1}{c|}{\begin{tabular}[c]{@{}c@{}}0.0120\\ (0.0248)\end{tabular}} &
\begin{tabular}[c]{@{}c@{}}0.0446\\ (0.0220)\end{tabular} & \begin{tabular}[c]{@{}c@{}}0.1575\\ (0.0798)\end{tabular} & \begin{tabular}[c]{@{}c@{}}0.0199\\ (0.0106)\end{tabular} \\
\hline
P=4 H=4 &
\begin{tabular}[c]{@{}c@{}}0.0188\\ (0.0288)\end{tabular} & \begin{tabular}[c]{@{}c@{}}0.0844\\ (0.1299)\end{tabular} & \multicolumn{1}{c|}{\begin{tabular}[c]{@{}c@{}}0.0098\\ (0.0185)\end{tabular}} &
\begin{tabular}[c]{@{}c@{}}0.0308\\ (0.0193)\end{tabular} & \begin{tabular}[c]{@{}c@{}}0.1018\\ (0.0851)\end{tabular} & \begin{tabular}[c]{@{}c@{}}0.0179\\ (0.0084)\end{tabular} \\
\hline
P=4 H=5 &
\begin{tabular}[c]{@{}c@{}}0.0172\\ (0.0272)\end{tabular} & \begin{tabular}[c]{@{}c@{}}0.0722\\ (0.1176)\end{tabular} & \multicolumn{1}{c|}{\begin{tabular}[c]{@{}c@{}}0.0104\\ (0.0198)\end{tabular}} &
\begin{tabular}[c]{@{}c@{}}0.0285\\ (0.0161)\end{tabular} & \begin{tabular}[c]{@{}c@{}}0.0942\\ (0.0769)\end{tabular} & \begin{tabular}[c]{@{}c@{}}0.0171\\ (0.0082)\end{tabular} \\
\hline
P=5 H=3 &
\begin{tabular}[c]{@{}c@{}}0.0255\\ (0.0273)\end{tabular} & \begin{tabular}[c]{@{}c@{}}0.1418\\ (0.1654)\end{tabular} & \multicolumn{1}{c|}{\begin{tabular}[c]{@{}c@{}}0.0095\\ (0.0174)\end{tabular}} &
\begin{tabular}[c]{@{}c@{}}0.0424\\ (0.0200)\end{tabular} & \begin{tabular}[c]{@{}c@{}}0.1499\\ (0.0762)\end{tabular} & \begin{tabular}[c]{@{}c@{}}0.0193\\ (0.0109)\end{tabular} \\
\hline
P=5 H=4 &
\begin{tabular}[c]{@{}c@{}}0.0159\\ (0.0227)\end{tabular} & \begin{tabular}[c]{@{}c@{}}0.0678\\ (0.1022)\end{tabular} & \multicolumn{1}{c|}{\begin{tabular}[c]{@{}c@{}}0.0100\\ (0.0168)\end{tabular}} &
\begin{tabular}[c]{@{}c@{}}0.0263\\ (0.0119)\end{tabular} & \begin{tabular}[c]{@{}c@{}}0.0832\\ (0.0485)\end{tabular} & \begin{tabular}[c]{@{}c@{}}0.0159\\ (0.0050)\end{tabular} \\
\hline
P=5 H=5 &
\begin{tabular}[c]{@{}c@{}}0.0188\\ (0.0272)\end{tabular} & \begin{tabular}[c]{@{}c@{}}0.0769\\ (0.1144)\end{tabular} & \multicolumn{1}{c|}{\begin{tabular}[c]{@{}c@{}}0.0102\\ (0.0174)\end{tabular}} &
\begin{tabular}[c]{@{}c@{}}0.0315\\ (0.0202)\end{tabular} & \begin{tabular}[c]{@{}c@{}}0.1067\\ (0.0927)\end{tabular} & \begin{tabular}[c]{@{}c@{}}0.0184\\ (0.0095)\end{tabular} \\
\hline
\multirow{3}{*}{SNR=2.5} &
\multicolumn{3}{c|}{TvReg} &
\multicolumn{3}{c|}{Mixed Graphical Models} \\
\cline{2-7}
\multicolumn{1}{|l|}{} & \multicolumn{3}{c|}{VAR Matrices {$A_{j,t}$}} & \multicolumn{3}{c|}{VAR  Matrices {$A_{j,t}$}} \\
\cline{2-7}
\multicolumn{1}{|l|}{} & All entries & True non-zeroes & \multicolumn{1}{l|}{True zeroes } & All entries & True non-zeroes  & \multicolumn{1}{l|}{True zeroes} \\
\hline
P=2 &
\begin{tabular}[c]{@{}c@{}}0.0922\\ (0.0106)\end{tabular} & \begin{tabular}[c]{@{}c@{}}0.2801\\ (0.0514)\end{tabular} & \multicolumn{1}{c|}{\begin{tabular}[c]{@{}c@{}}0.0610\\ (0.0067)\end{tabular}} &
\begin{tabular}[c]{@{}c@{}}0.1150\\ (0.0160)\end{tabular} & \begin{tabular}[c]{@{}c@{}}0.4479\\ (0.0386)\end{tabular} & \begin{tabular}[c]{@{}c@{}}0.0124\\ (0.0039)\end{tabular} \\
\hline
P=3 &
\begin{tabular}[c]{@{}c@{}}0.1070\\ (0.0116)\end{tabular} & \begin{tabular}[c]{@{}c@{}}0.3205\\ (0.0497)\end{tabular} & \multicolumn{1}{c|}{\begin{tabular}[c]{@{}c@{}}0.0718\\ (0.0081)\end{tabular}} &
\begin{tabular}[c]{@{}c@{}}0.1165\\ (0.0161)\end{tabular} & \begin{tabular}[c]{@{}c@{}}0.4539\\ (0.0374)\end{tabular} & \begin{tabular}[c]{@{}c@{}}0.0121\\ (0.0039)\end{tabular} \\
\hline
P=4 &
\begin{tabular}[c]{@{}c@{}}0.1163\\ (0.0108)\end{tabular} & \begin{tabular}[c]{@{}c@{}}0.3328\\ (0.0456)\end{tabular} & \multicolumn{1}{c|}{\begin{tabular}[c]{@{}c@{}}0.0825\\ (0.0073)\end{tabular}} &
\begin{tabular}[c]{@{}c@{}}0.1170\\ (0.0162)\end{tabular} & \begin{tabular}[c]{@{}c@{}}0.4560\\ (0.0374)\end{tabular} & \begin{tabular}[c]{@{}c@{}}0.0118\\ (0.0033)\end{tabular} \\
\hline
P=5 &
\begin{tabular}[c]{@{}c@{}}0.1255\\ (0.0111)\end{tabular} & \begin{tabular}[c]{@{}c@{}}0.3402\\ (0.0444)\end{tabular} & \multicolumn{1}{c|}{\begin{tabular}[c]{@{}c@{}}0.0940\\ (0.0084)\end{tabular}} &
\begin{tabular}[c]{@{}c@{}}0.1172\\ (0.0161)\end{tabular} & \begin{tabular}[c]{@{}c@{}}0.4571\\ (0.0374)\end{tabular} & \begin{tabular}[c]{@{}c@{}}0.0114\\ (0.0036)\end{tabular} \\
\hline
\end{tabular}
}
\end{table}

\begin{table}
\caption{\label{tab:gamma-performance-2} Simulation study 1; SNR = 2.5. Performance evaluation of the posterior estimation of the components' indicators $(\gamma_{h,P+1},\dots,\gamma_{h,T})$, based on average accuracy, sensitivity, specificity and precision across the 30 generated data sets. Standard deviations are indicated in parentheses. See Section 6 for details.\\}
\centering
\resizebox{\textwidth}{0.15\textheight}{
\begin{tabular}{|c|cccc|c|cccc|}
\hline
\multirow{2}{*}{SNR=2.5} & \multicolumn{4}{c|}{$(\gamma_{h,P+1},\dots,\gamma_{h,T})$} & \multirow{2}{*}{SNR=2.5} & \multicolumn{4}{c|}{$(\gamma_{h,P+1},\dots,\gamma_{h,T})$}  \\ \cline{2-5} \cline{7-10}
& \multicolumn{1}{c}{Acc} & \multicolumn{1}{c}{Sens} & \multicolumn{1}{c}{Spec} & \multicolumn{1}{c|}{Prec} & & \multicolumn{1}{c}{Acc} & \multicolumn{1}{c}{Sens} & \multicolumn{1}{c}{Spec} & \multicolumn{1}{c|}{Prec} \\
\hline
P=2 H=3 &
\begin{tabular}[c]{@{}c@{}}0.8669\\ (0.2110)\end{tabular} & \begin{tabular}[c]{@{}c@{}}0.7069\\ (0.4343)\end{tabular} & \begin{tabular}[c]{@{}c@{}}0.9812\\ (0.0933)\end{tabular} & \begin{tabular}[c]{@{}c@{}}0.9579\\ (0.1499)\end{tabular} &
P=4 H=3 &
\begin{tabular}[c]{@{}c@{}}0.8696\\ (0.2236)\end{tabular} & \begin{tabular}[c]{@{}c@{}}0.7160\\ (0.4450)\end{tabular} & \begin{tabular}[c]{@{}c@{}}0.9826\\ (0.0938)\end{tabular} & \begin{tabular}[c]{@{}c@{}}0.9672\\ (0.1465)\end{tabular} \\
\hline
P=2 H=4 &
\begin{tabular}[c]{@{}c@{}}0.9214\\ (0.1750)\end{tabular} & \begin{tabular}[c]{@{}c@{}}0.8671\\ (0.3096)\end{tabular} & \begin{tabular}[c]{@{}c@{}}0.9353\\ (0.2137)\end{tabular} & \begin{tabular}[c]{@{}c@{}}0.9308\\ (0.2069)\end{tabular} &
P=4 H=4 &
\begin{tabular}[c]{@{}c@{}}0.9464\\ (0.1453)\end{tabular} & \begin{tabular}[c]{@{}c@{}}0.8740\\ (0.3192)\end{tabular} & \begin{tabular}[c]{@{}c@{}}0.9884\\ (0.0930)\end{tabular} & \begin{tabular}[c]{@{}c@{}}0.9716\\ (0.1539)\end{tabular} \\
\hline
P=2 H=5 &
\begin{tabular}[c]{@{}c@{}}0.9276\\ (0.1605)\end{tabular} & \begin{tabular}[c]{@{}c@{}}0.8698\\ (0.3195)\end{tabular} & \begin{tabular}[c]{@{}c@{}}0.9437\\ (0.1930)\end{tabular} & \begin{tabular}[c]{@{}c@{}}0.9370\\ (0.1948)\end{tabular} &
P=4 H=5 &
\begin{tabular}[c]{@{}c@{}}0.9516\\ (0.1355)\end{tabular} & \begin{tabular}[c]{@{}c@{}}0.9064\\ (0.2769)\end{tabular} & \begin{tabular}[c]{@{}c@{}}0.9679\\ (0.1438)\end{tabular} & \begin{tabular}[c]{@{}c@{}}0.9623\\ (0.1536)\end{tabular} \\
\hline
P=3 H=3 &
\begin{tabular}[c]{@{}c@{}}0.8731\\ (0.2123)\end{tabular} & \begin{tabular}[c]{@{}c@{}}0.7198\\ (0.4451)\end{tabular} & \begin{tabular}[c]{@{}c@{}}0.9751\\ (0.1373)\end{tabular} & \begin{tabular}[c]{@{}c@{}}0.9722\\ (0.1322)\end{tabular} &
P=5 H=3 &
\begin{tabular}[c]{@{}c@{}}0.8704\\ (0.2319)\end{tabular} & \begin{tabular}[c]{@{}c@{}}0.7271\\ (0.4398)\end{tabular} & \begin{tabular}[c]{@{}c@{}}0.9926\\ (0.0330)\end{tabular} & \begin{tabular}[c]{@{}c@{}}0.9761\\ (0.1241)\end{tabular} \\
\hline
P=3 H=4 &
\begin{tabular}[c]{@{}c@{}}0.9549\\ (0.1227)\end{tabular} & \begin{tabular}[c]{@{}c@{}}0.8874\\ (0.3066)\end{tabular} & \begin{tabular}[c]{@{}c@{}}0.9783\\ (0.1329)\end{tabular} & \begin{tabular}[c]{@{}c@{}}0.9813\\ (0.0869)\end{tabular} &
P=5 H=4 &
\begin{tabular}[c]{@{}c@{}}0.9566\\ (0.1396)\end{tabular} & \begin{tabular}[c]{@{}c@{}}0.9237\\ (0.2512)\end{tabular} & \begin{tabular}[c]{@{}c@{}}0.9765\\ (0.1381)\end{tabular} & \begin{tabular}[c]{@{}c@{}}0.9735\\ (0.1357)\end{tabular} \\
\hline
P=3 H=5 &
\begin{tabular}[c]{@{}c@{}}0.9704\\ (0.1025)\end{tabular} & \begin{tabular}[c]{@{}c@{}}0.9276\\ (0.2502)\end{tabular} & \begin{tabular}[c]{@{}c@{}}0.9856\\ (0.0776)\end{tabular} & \begin{tabular}[c]{@{}c@{}}0.9697\\ (0.1585)\end{tabular} &
P=5 H=5 &
\begin{tabular}[c]{@{}c@{}}0.9548\\ (0.1331)\end{tabular} & \begin{tabular}[c]{@{}c@{}}0.9018\\ (0.2889)\end{tabular} & \begin{tabular}[c]{@{}c@{}}0.9732\\ (0.1415)\end{tabular} & \begin{tabular}[c]{@{}c@{}}0.9621\\ (0.1724)\end{tabular} \\
\hline
\end{tabular}
}
\end{table}

\begin{table}
\caption{\label{tab:components-performance-3} Simulation study 2. Posterior means of estimated tensor components $A^*_{j,h}$ and  coefficients $A_{j,t}$ from the BTVT-VAR model, compared with frequentist estimates from the \textbf{tvReg} and \textbf{mgm} R-packages. Tensor components are evaluated using the square root of the average Frobenius norm of the difference between the posterior mean and true components. Columns 2 and 3 show average Euclidean distances for truly non-zero and zero entries, with standard deviations in parentheses. See Section 6 for details.\\}
%Bayesian point estimates (posterior means) of the identified tensor components and dynamic coefficients from the proposed BTVT-VAR model. The latter are compared with the frequentist estimates of a time-varying VAR model implemented in the \textbf{tvReg} and the \textbf{mgm} R-package. 
%The evaluation of the tensor components is based on the  square-root of the average Frobenius norm of the difference  between the posterior mean and the true matrices across three true components. Columns 2 and 3 show the average Euclidean distances for each  truly non-zero and truly zero entry in the matrices. Standard deviations across the three true components are indicated in parentheses. See Section 6 for details.}
\centering
\resizebox{!}{!}{
\begin{tabular}{|l|llll}
\hline
\multicolumn{1}{|l}{} &
\multicolumn{1}{l}{} &
\multicolumn{1}{c}{All entries} & \multicolumn{1}{c}{True non-zeroes} & \multicolumn{1}{c|}{True zeroes} \\
\hline
\multicolumn{1}{|c|}{\multirow{2}{*}{BTVT-VAR}} &
\multicolumn{1}{c} \text{Latent} {$A_{j,h}^\ast$} & \multicolumn{1}{|c}{\begin{tabular}[c]{@{}c@{}}\multicolumn{1}{c}{0.0141}\\ \multicolumn{1}{c}{(0.0016)}\end{tabular}} &
\multicolumn{1}{c}{\begin{tabular}[c]{@{}c@{}}\multicolumn{1}{c}{0.0504}\\ \multicolumn{1}{c}{(0.0369)}\end{tabular}} & \multicolumn{1}{c|}{\begin{tabular}[c]{@{}c@{}}\multicolumn{1}{c}{0.0114}\\ \multicolumn{1}{c}{(0.0032)}\end{tabular}} \\
\cline{2-5}
&
\multicolumn{1}{c} \text{VAR Matrices }{$A_{j,t}$} & \multicolumn{1}{|c}{\begin{tabular}[c]{@{}c@{}}\multicolumn{1}{c}{0.0188}\end{tabular}} & \multicolumn{1}{c}{\begin{tabular}[c]{@{}c@{}}\multicolumn{1}{c}{0.0365}\end{tabular}} & \multicolumn{1}{c|}{\begin{tabular}[c]{@{}c@{}}\multicolumn{1}{c}{0.0158}\end{tabular}} \\
\hline
\multicolumn{1}{|c}{TvReg} &
\multicolumn{1}{|l} \text{VAR Matrices }{$A_{j,t}$} & \multicolumn{1}{|c}{\begin{tabular}[c]{@{}c@{}}\multicolumn{1}{c}{0.3798}\end{tabular}} & \multicolumn{1}{c}{\begin{tabular}[c]{@{}c@{}}\multicolumn{1}{c}{0.2717}\end{tabular}} & \multicolumn{1}{c|}{\begin{tabular}[c]{@{}c@{}}\multicolumn{1}{c}{0.3895}\end{tabular}} \\
\hline
\multicolumn{1}{|c}{MGM} &
\multicolumn{1}{|l} \text{VAR  Matrices }{$A_{j,t}$} & \multicolumn{1}{|c}{\begin{tabular}[c]{@{}c@{}}\multicolumn{1}{c}{0.1072}\end{tabular}} & \multicolumn{1}{c}{\begin{tabular}[c]{@{}c@{}}\multicolumn{1}{c}{0.3459}\end{tabular}} & \multicolumn{1}{c|}{\begin{tabular}[c]{@{}c@{}}\multicolumn{1}{c}{0.0090}\end{tabular}} \\
\hline
\end{tabular}
}
\end{table}

\begin{table}
\caption{\label{tab:gamma-performance-3} Simulation study 2. Performance evaluation of the posterior estimation of the components' indicators $(\gamma_{h,P+1},\dots,\gamma_{h,T})$. The evaluation is based on average accuracy, sensitivity, specificity and precision over the true components.\\}
\centering
\resizebox{!}{!}{
\begin{tabular}{lllll}
\hline
\multicolumn{1}{|l}{} & \multicolumn{1}{c}{Acc} & \multicolumn{1}{c}{Sens} & \multicolumn{1}{c}{Spec} & \multicolumn{1}{c|}{Prec} \\
\hline
\multicolumn{1}{|l|}{$(\gamma_{h,P+1},\dots,\gamma_{h,T})$} & \multicolumn{1}{c}{\begin{tabular}[c]{@{}c@{}}\multicolumn{1}{c}{0.9955}\\ \multicolumn{1}{c}{(0.0078)}\end{tabular}} & \multicolumn{1}{c}{\begin{tabular}[c]{@{}c@{}}\multicolumn{1}{c}{1.0000}\\ \multicolumn{1}{c}{(0.0000)}\end{tabular}} & \multicolumn{1}{c}{\begin{tabular}[c]{@{}c@{}}\multicolumn{1}{c}{0.9911}\\ \multicolumn{1}{c}{ (0.0154)}\end{tabular}} &\multicolumn{1}{c|}{\begin{tabular}[c]{@{}c@{}}\multicolumn{1}{c}{0.9911}\\ \multicolumn{1}{c}{ (0.0154)}\end{tabular}}\\ \hline
\end{tabular}
}
\end{table}

\begin{figure}
     \centering
     \begin{subfigure}[b]{0.3\textwidth}
         \centering
         \includegraphics[height=0.4\textheight]{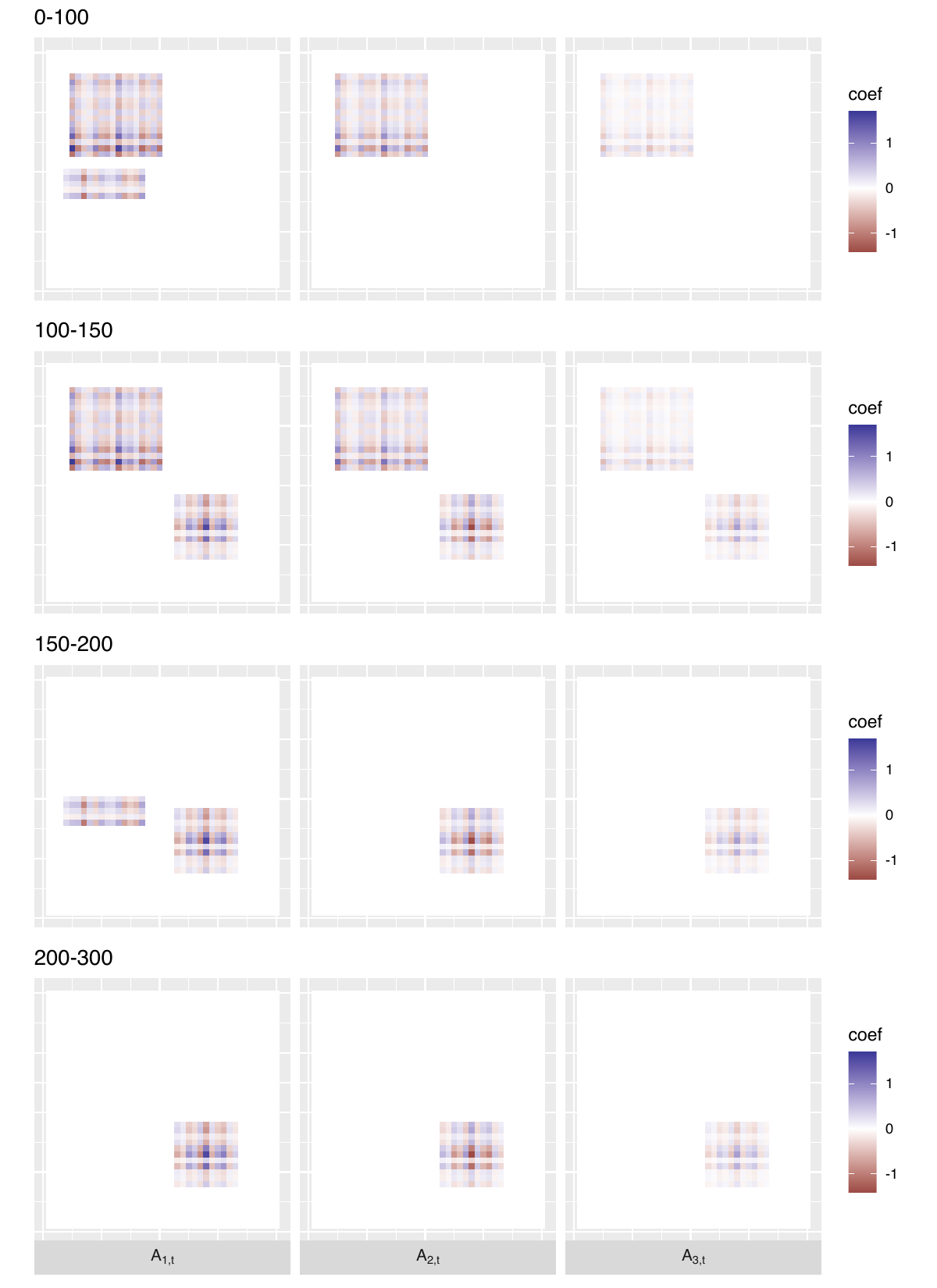}
         \caption{Truth}
     \end{subfigure}
     \hfill
     \begin{subfigure}[b]{0.6\textwidth}
         \centering
         \includegraphics[height=0.4\textheight]{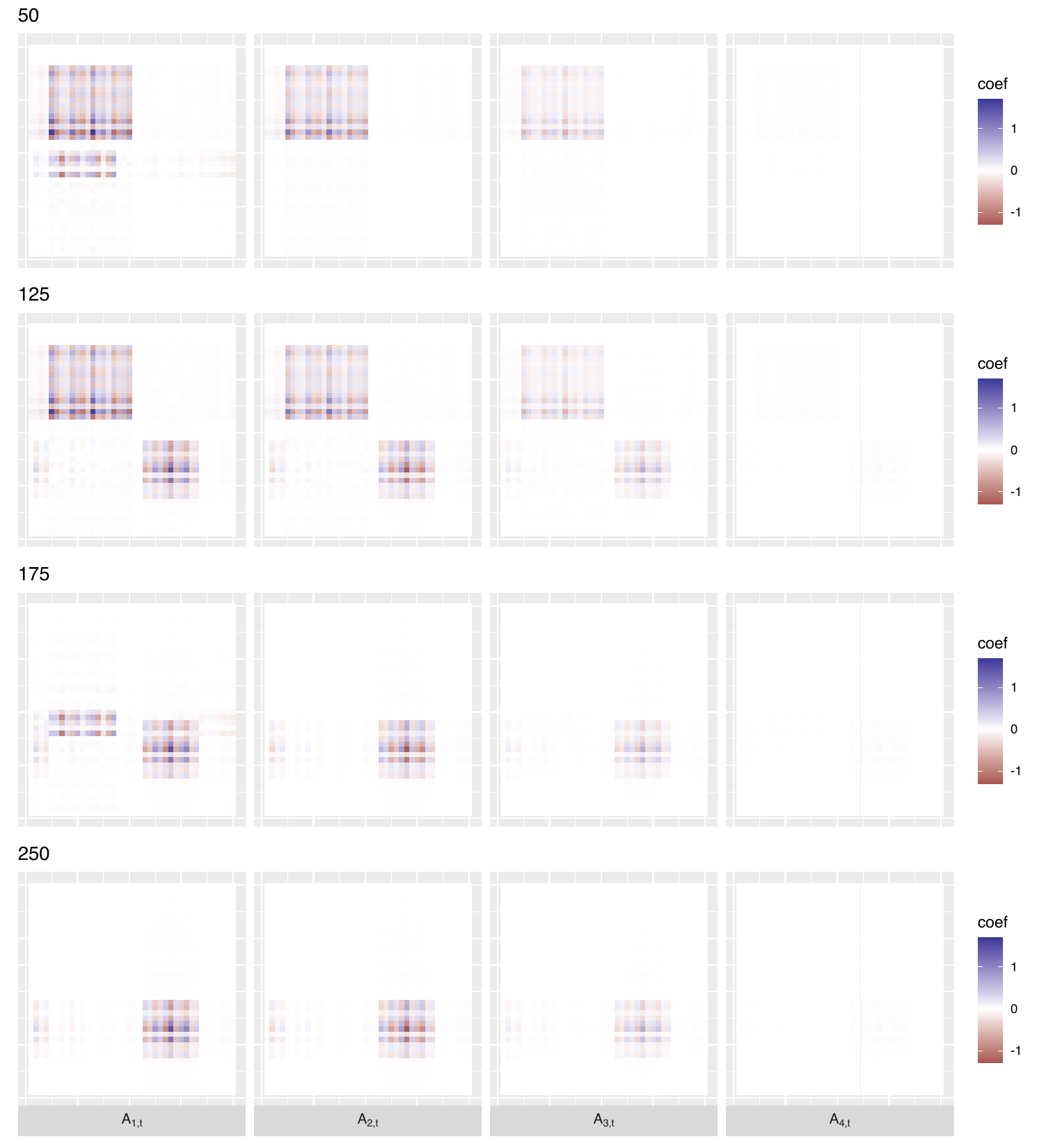}
         \caption{$E(A_{p,t}|y)$}
     \end{subfigure}
        \caption{Comparison between the true BTVT-VAR coefficients $A_{1,t}, A_{2,t}, A_{3,t}$ on the left panel and the MCMC  posterior means from the BTV-TVAR model on the right panel in the second simulation study of Section \ref{Simulation Studies}. For each truly invariant time window, the estimated matrices at time point 50, 125, 175 and 250 are displayed.}
        \label{fig:simulation 3}
\end{figure}

\begin{figure}
    \centering
    \includegraphics[width=\linewidth]{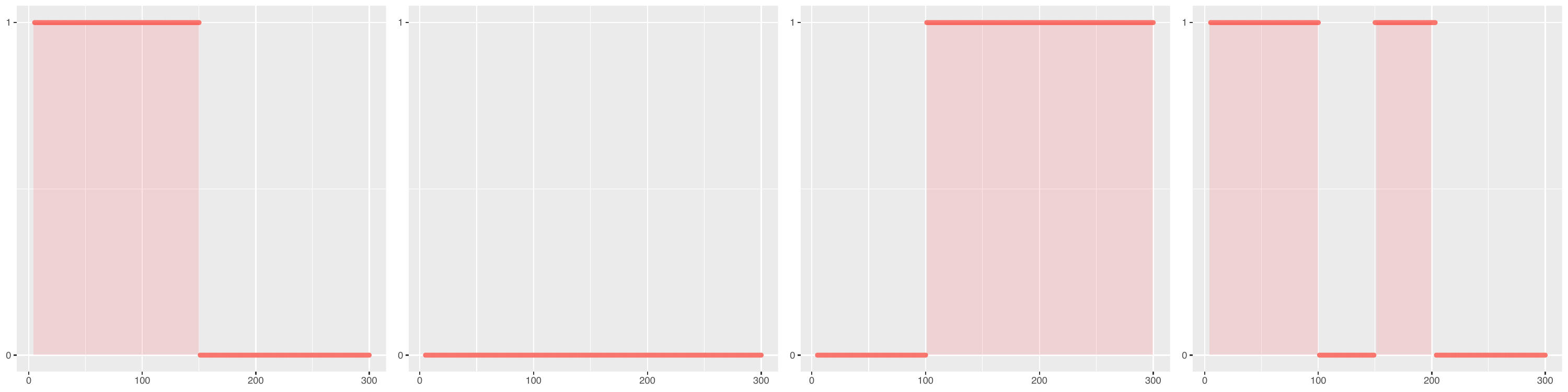}
    \caption{Estimated trajectories of the latent indicators of activation $\gamma_{h,t}$ for four components $h=1, \ldots, 4$. The red areas indicate the true activations. The solid lines indicates the estimated values of $\gamma_{h,t}$ based on posterior probabilities of activation greater than $0.5$.}  
    \label{fig:simulation 3 traj}
\end{figure}

\emph{Simulation 2.} We consider a time series from a larger $N=40$ TV-VAR model of order $P=3$ with dynamic coefficients shown in the left panel of Figure \ref{fig:simulation 3}. These coefficients are combinations of $H=3$ components. A total of $T=300$ observations are simulated; %among which 
the coefficients matrix of the first 200 observations admit a rank-2 tensor decomposition with time varying %changing 
mixing components whereas the last 100 observations consist of only one component. The covariance matrix $\Sigma$ of the error term $\epsilon_t$ has diagonal elements %$c((1:25)/5, (25:11)/5)$. 
$\Sigma[i,i]=i/5$ for $i=1, \ldots, 25$ and $\Sigma[i,i]=(51-k)/5$ for $i=26, \ldots, 40$. %The goal of this simulation study is to validate the performance of the ($\text{MC}^3$) algorithm \bchm Not sure why we say this, since we do not comment on th algorithm \ech. %, so in the 
For the posterior inference, we run the algorithm for 9,000 iterations, with swap being performed every 300 iterations. For more details on the algorithm, refer to Section \ref{Real Data Application} and the supplementary material \citep{zhang2024}. We choose $H=4$ and $P=4$. Even though we include one more component and lag, we expect the extra component as well as the coefficients at lag 4 to be close to zero due to the effect of the shrinkage priors. The remaining experiment settings are the same as in the first simulation. 
We select four estimated coefficients matrices at time point 50, 125, 175 and 250 as in the right panel of Figure \ref{fig:simulation 3}. Our method identifies three ``non-empty'' components and it accurately captures the patterns of the true ones. Furthermore, the dynamics of the coefficient matrices are all accurately identified by the model. Table \ref{tab:components-performance-3} shows an evaluation of the posterior inference on the coefficient matrices in our model versus a frequentist time-varying regression. Once again, our model compares quite favorably. To further verify our method's ability to detect changing patterns along all the 296 time points, we report the accuracy, sensitivity, specificity and precision in the estimation of the latent activation indicators $\gamma_{h,t}$'s in Table \ref{tab:gamma-performance-3}.  In Figure \ref{fig:simulation 3 traj} we report the estimated  trajectories of the $\gamma_{h,t}$ along %as a function of 
time, for each component $h=1, \ldots, H$. The shaded red areas indicate the true component activation, whereas the solid line indicates the posterior modes. 
One of the trajectories, being constantly zero, identifies an empty component: indeed, we employed $H=4$ for model fitting instead of the true number of components. The other three estimated trajectories follow the true activations quite closely, reaching false positive rates of 0\%, 2.67\% and 0\% as well as false negative rate of 0\%, 0\% and 0\%, respectively.
Once again, the results illustrate the role of the shrinkage prior specifications, since fixing higher values of $H$ and $P$  does not appear to hamper the estimation of the VAR matrices. In particular, we do not need to rely on model selection techniques in order to determine the appropriate values of $H$ and $P$. Therefore,  in many cases it may be desirable to learn the actual dimensions of the model from the data, by fixing relatively large values of $H$ and $P$.\\
As a final remark, we note that we also considered a simulation scenario with $N=100$ multivariate time series. However, the frequentist TvReg approach did show numerical problems with such a large number of times series, after taking 9.44 hours to complete. On the contrary, our method was still able to obtain good inferences for this large dimension case, after taking 3.6 hours to complete 5,000 iterations on a Intel Core i5-6300U CPU at 2.40GHz, with 8GB RAM. For comparison, using our method, each run of the $N=40$-dimensional case took approximately 56 minutes to complete.

\section{Time-varying Effective Connectivity in an fMRI reading experiment}
\label{Real Data Application}
We apply our BTVT-VAR model to a task-based functional magnetic resonance imaging (fMRI) data set \citep{wehbe}. The experiment involved participants aged 18 to 40, who were tasked with reading Chapter 9 of \textit{Harry Potter and the Sorcerer’s Stone} \citep{Rowling2012}.  All subjects had previously read the book or seen the movie. The words of the story were presented in rapid succession, where each word was presented one by one at the center of the screen for 0.5 seconds in black font on a gray background. A Siemens Verio 3.0T scanner was used to acquire the scans, utilizing a T2$^{*}$ sensitive echo planar imaging pulse sequence with repetition time (TR) of 2s, time echo (TE) of 29 ms, flip angle (FA) of 79$^{\circ}$, 36 number of slices and $3\times3\times3$mm$^3$ voxels. Data preprocessing involved realignment, slice timing correction, and co-registration with the subject’s anatomical scan, which was segmented into grey and white matter and cerebro-spinal fluid. The subjects' scans were then normalized to the Montreal Neurological Institute (MNI) template and smoothed with a 6 $\times$ 6 $\times$ 6mm Gaussian kernel smoother.  For more details, see \cite{ondrus}.

\begin{figure}
     \centering
     \begin{subfigure}[b]{0.47\textwidth}
         \centering
         \includegraphics[width=\textwidth]{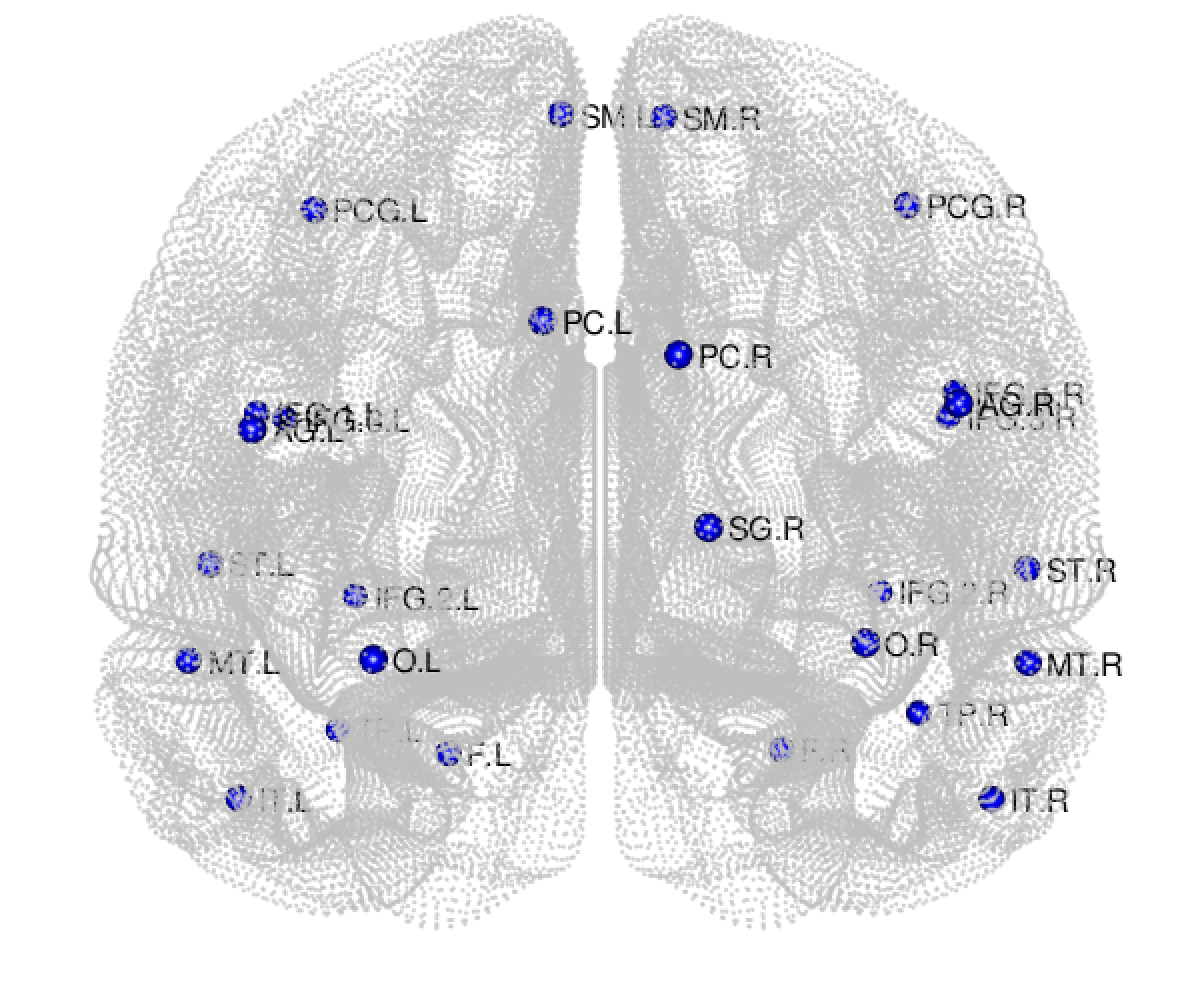}
         \caption{}
     \end{subfigure}
     \begin{subfigure}[b]{0.47\textwidth}
         \centering
         \includegraphics[width=\textwidth]{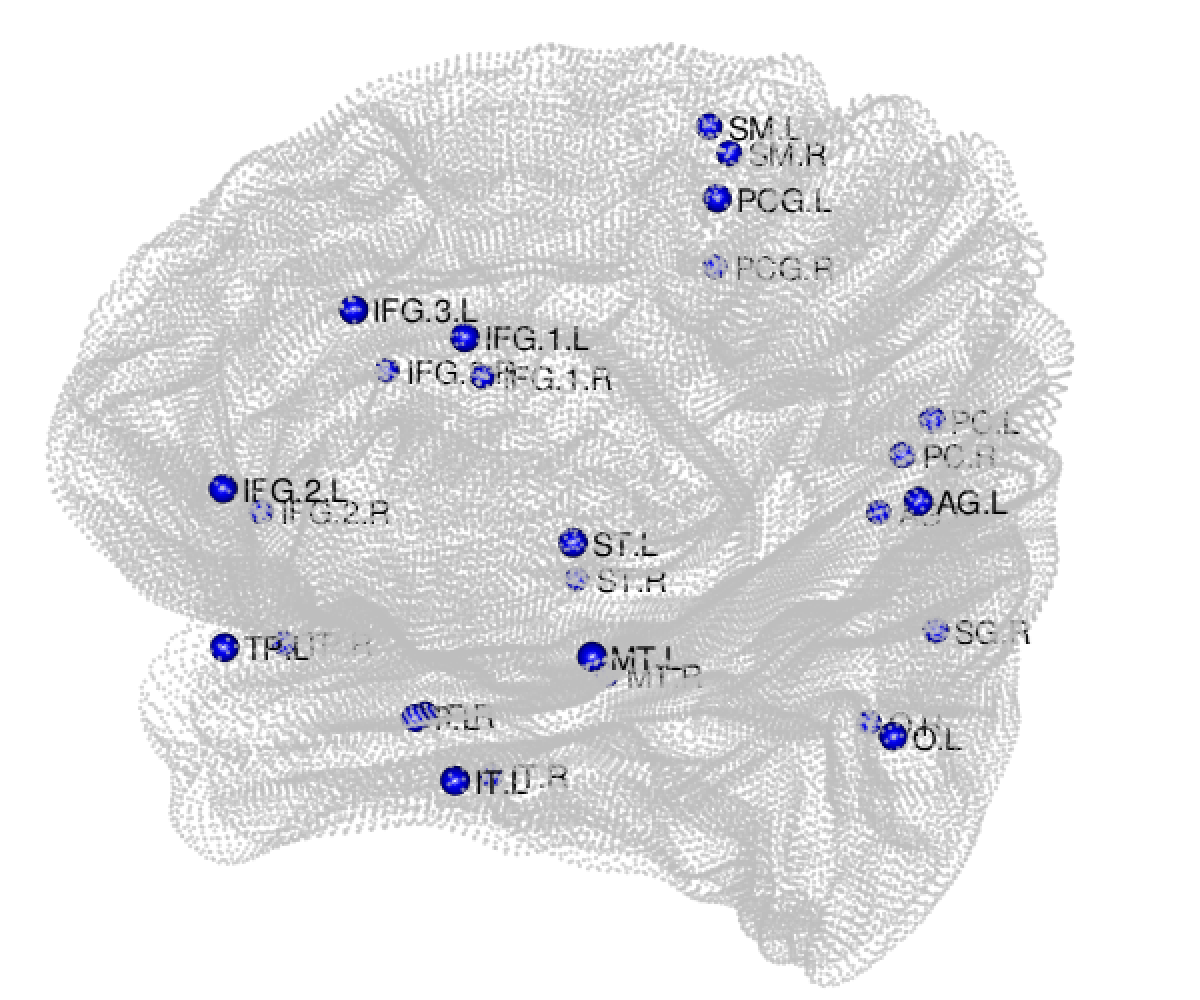}
         \caption{}
     \end{subfigure}
     \caption{Locations of the 27 ROIs extracted from the Harry Potter task-based fMRI data set. Each ROI has a left and right hemisphere component apart from the supramarginal gyrus region.\\}
     \label{fig:ROIs}
\end{figure}

Twenty-seven ROIs defined using the Automated Anatomical Atlas (AAL) brain atlas \citep{tzourio} were extracted from the data set, shown in Table \ref{HPROI}. Figure \ref{fig:ROIs}, created using R package \textbf{fabisearch} by \citet{ondrus_package,ondrus2024}, shows the locations of the ROIs.  These regions contain a variety of voxels that have been previously recognized as important to distinguish between the literary content of two novel text passages based on neural activity while these passages are being read \citep{Xiong}. 

We expect that the dynamics of effective connectivity inferred through our BTVT-VAR model coincide with different events in the plot.  To illustrate, we focus on data from a representative subject and on a time point which includes a thrilling moment in the story: Harry, Ron, and Hermione (the main three characters in the book) arrive in a forbidden corridor, turn around, and come face-to-face with a monstrous three-headed dog. This occurs at approximately time point $t=1176$ in the experiment.  While we are in possession of the event information, our model does not utilize such information.  For model fitting, we set $H=8$, allowing us to capture over 50\% of the estimated variability in the sample, as determined by the Frobenious norm of the coefficient matrices estimated in a frequentist multivariate autoregressive integrated moving average (MARIMA) model. Additionally, we set $P=4$. In applications of VAR models to fMRI data, it's common to consider only an $AR(1)$ structure, which is seen as an effective representation of short-range temporal dependencies over a small number of ROIs, typically fewer than ten.

\begin{table}
  \caption{Information on the 27 ROIs extracted from the Harry Potter task-based fMRI data set.  Each ROI has a left and right hemisphere component apart from the supramarginal gyrus region.\\}
    \centering
    \begin{tabular}{clcc}
    \toprule
    \textbf{ROI id} & \multicolumn{1}{c}{\textbf{Regions}}  & \textbf{Label} \\
    \midrule
    1 & Angular gyrus & AG \\
    \midrule
    2 & Fusiform gyrus & F \\
    \midrule
    3 & Inferior temporal gyrus & IT \\
    \midrule
    4 & Inferior frontal gyrus, opercular part & IFG 1 \\
    \midrule
    5 & Inferior frontal gyrus, orbital part & IFG 2 \\
    \midrule
    6 & Inferior frontal gyrus, triangular part & IFG 3 \\
    \midrule
    7 & Middle temporal gyrus & MT \\
    \midrule
    8 & Occipital lobe & O \\
    \midrule
    9 & Precental gyrus & PCG \\
    \midrule
    10 & Precuneus & PC \\
    \midrule
    11 & Supplementary motor area & SM \\
    \midrule
    12 & Superior temporal gyrus & ST \\
    \midrule
    13 & Temporal pole & TP \\
    \midrule
    14 & Supramarginal gyrus & SG.R \\
    \bottomrule
    \end{tabular}
  \label{HPROI}
\end{table}

\begin{figure}
\centering
\subfloat[]{
    \includegraphics[width=0.47\linewidth]{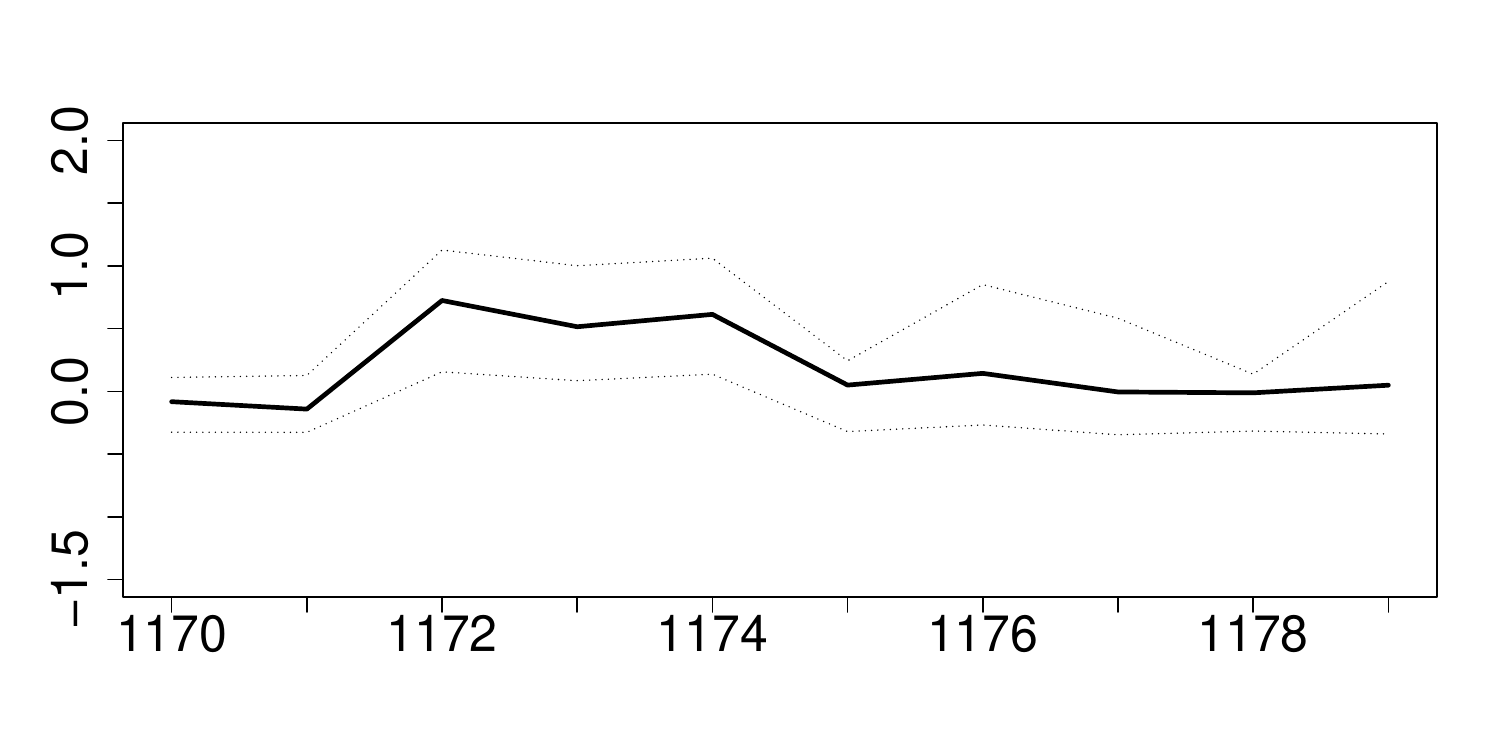}}
\subfloat[]{
    \includegraphics[width=0.47\linewidth]{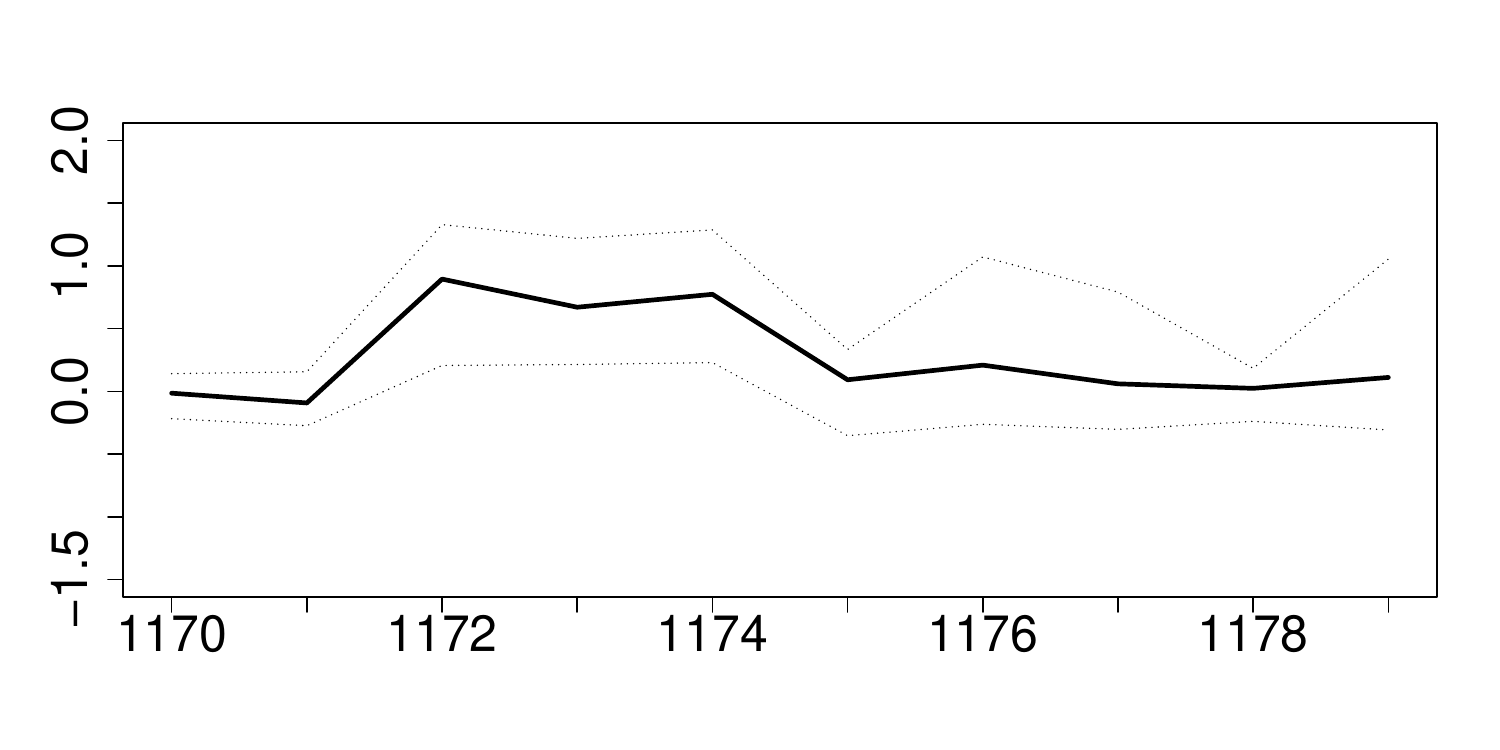}
}
\caption{The trajectories of selected coefficients (from F.R to ST.L on the left and from F.R to ST.R on the right) with 80\% credible intervals, for subject 4.}
\label{fig:sub4_traj2}
\end{figure}

 First, we challenge the model to identify possible changes in the effective connectivity that we are not aware of in advance and connect them to events in the story plot.  As an example, Figure \ref{fig:sub4_traj2} shows the estimated trajectories of selected coefficients (from ROI F.R to ROI ST.L on the left and from ROI F.R to ROI ST.R on the right) across several time points, with 80\% credible intervals, of the representative subject.  As explained previously, an important advantage of the BTVT-VAR model is its ability to estimate dynamic effective connectivity at each time point. The two coefficients exhibit similar time-varying patterns: they show large positive values before the encounter with the three-headed monsters and become less active thereafter. Despite the slight change in value, there is still an evident increase in dynamic effective connectivity between the ROIs near the pivotal plot event at time point $t=1176$. After estimating the $\hat{A}_t$ coefficients, it is possible to calculate the Frobenius distance $\Delta A_t= || A_t - A_{t-1}||_F$ between $A_t$ and $A_{t-1}$ as an indicator of the change in effective connectivity at time $t$. Section S6 in the Supplementary material presents these distances for the selected subject and the other participants in the study, emphasizing the changes in time-varying coefficients, particularly in the vicinity of the pivotal point of the story \citep{zhang2024}. Section S7 in the supplementary material further investigates the behavior of the lag-1 mean coefficients for subjects who exhibit more pronounced shifts in their brain dynamic effective connectivity patterns before and after the time point $t=1176$ \citep{zhang2024}. 

In the remainder of this manuscript, we explore the estimation of a Granger causality network, based on the estimated VAR coefficient matrices. Granger causality is a mainstream tool to infer and understand effective brain connectivity \citep{friston2013analysing}. Informally, Granger causality states that if including past values of $y_{i,t}$, the $i$th entry of the observed vector $\mathbf{y}_t$, improves the prediction of $y_{j,t}$, with respect to only including past values of $y_{j,t}$, then the $i$th entry Granger causes the $j$th entry at time $t$ \citep{granger1969investigating}. It is straightforward to infer Granger causality from the elements of the VAR coefficient matrices: $A_{p,t}[i,j]=0$ for all $p=1,\dots,P$ if and only if, at time $t$, the $i$-th entry does not cause the $j$-th entry in a Granger sense. In order to estimate the Granger causalities at each time point, we  estimate  the probability that the $i$th ROI does not Granger cause the $j$th ROI by computing
\begin{equation*}
    p_{j,i}(t) \equiv P\left(|A_{p,t}[j,i]|<\delta \text{ for all }p=1,\dots,P \mid \mathbf{y}_{1:T} \right)
\end{equation*}
for some small $\delta$. This probability can be easily approximated using the MCMC samples.

\begin{figure}
\centering
\subfloat[\label{subfig:GC_sub4baseline_0.01_p1}]{
    \includegraphics[width=0.32\linewidth]{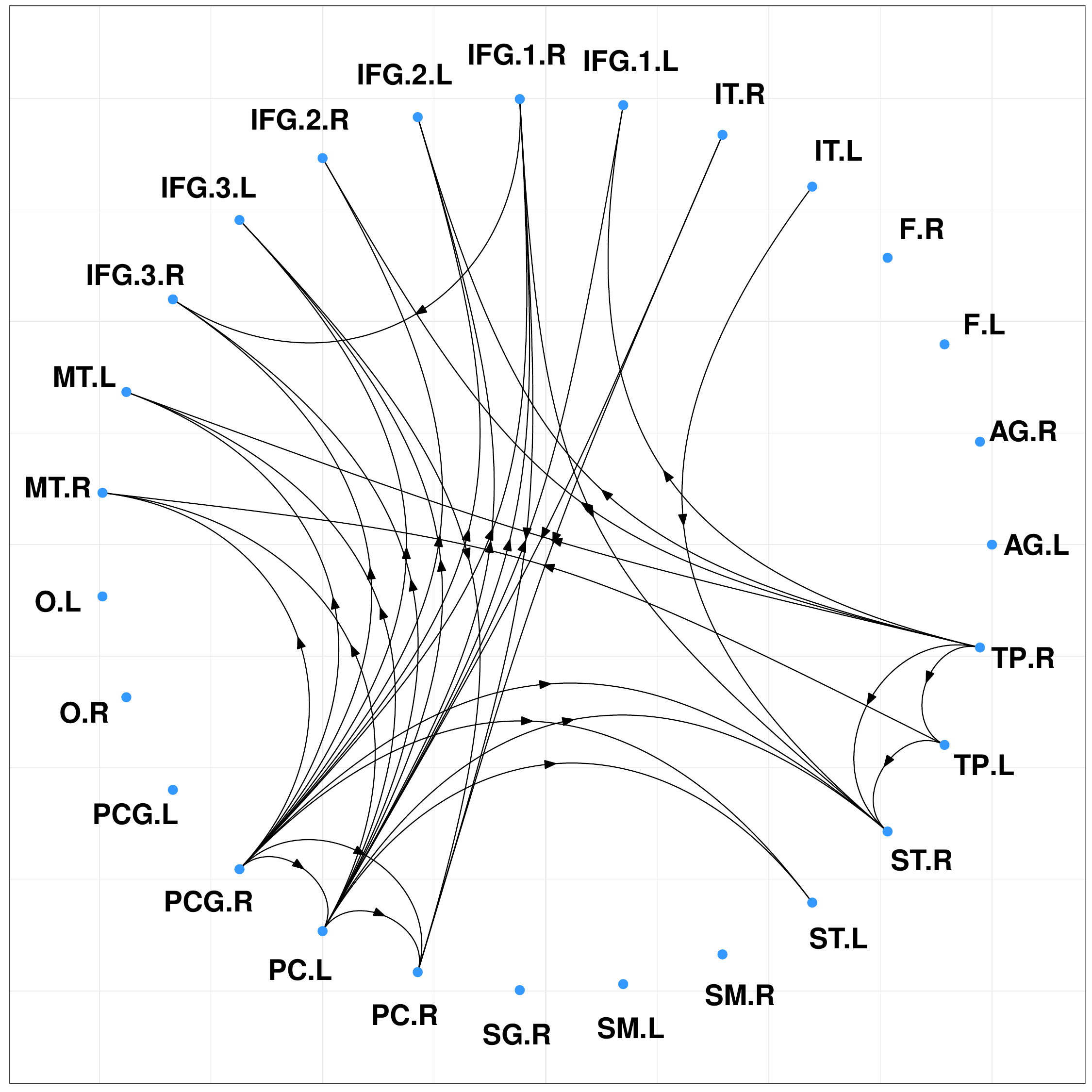}
}
\subfloat[\label{subfig:GC_sub4before_0.01_p1}]{
    \includegraphics[width=0.32\linewidth]{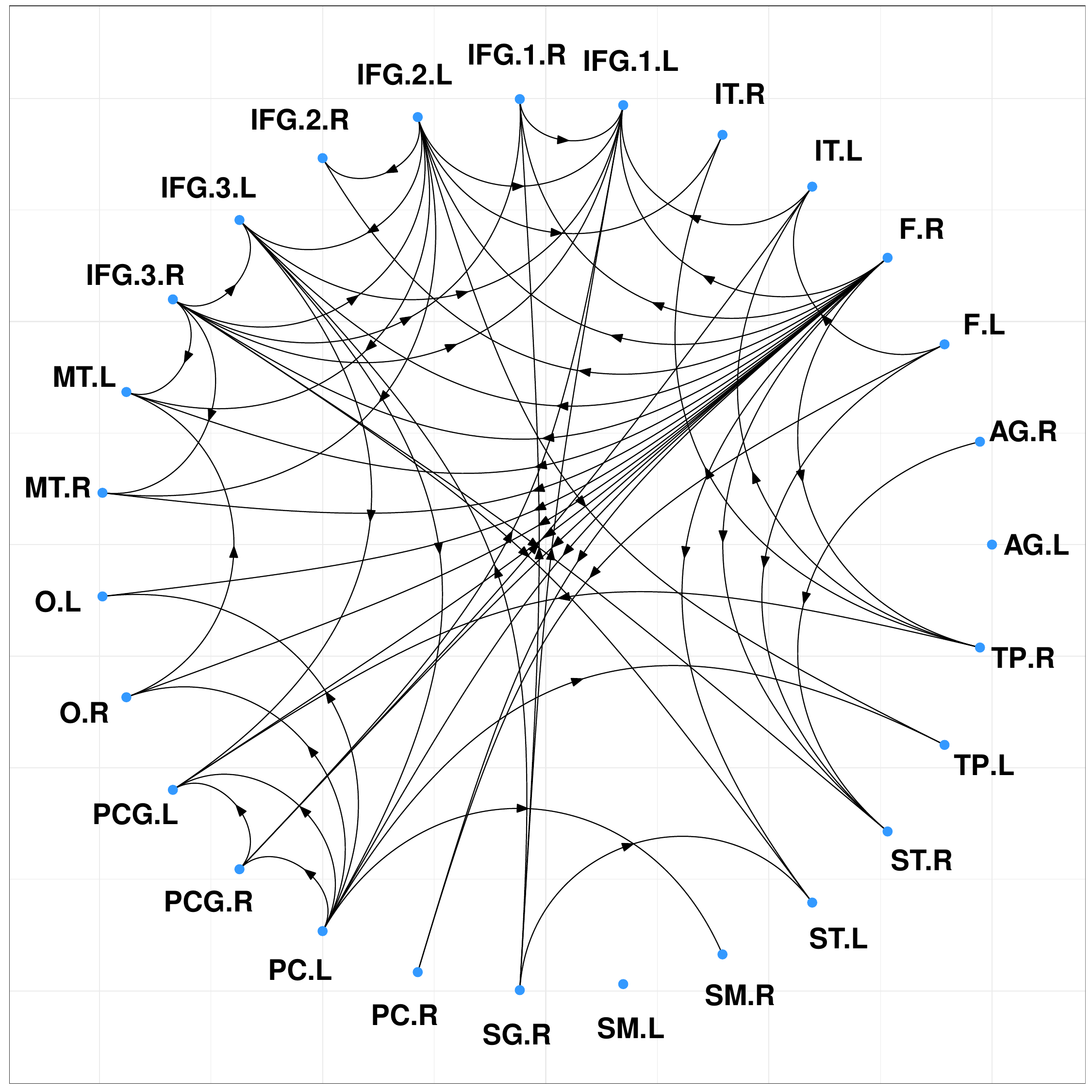}
}
\subfloat[\label{subfig:GC_sub4after_0.01_p1}]{
    \includegraphics[width=0.32\linewidth]{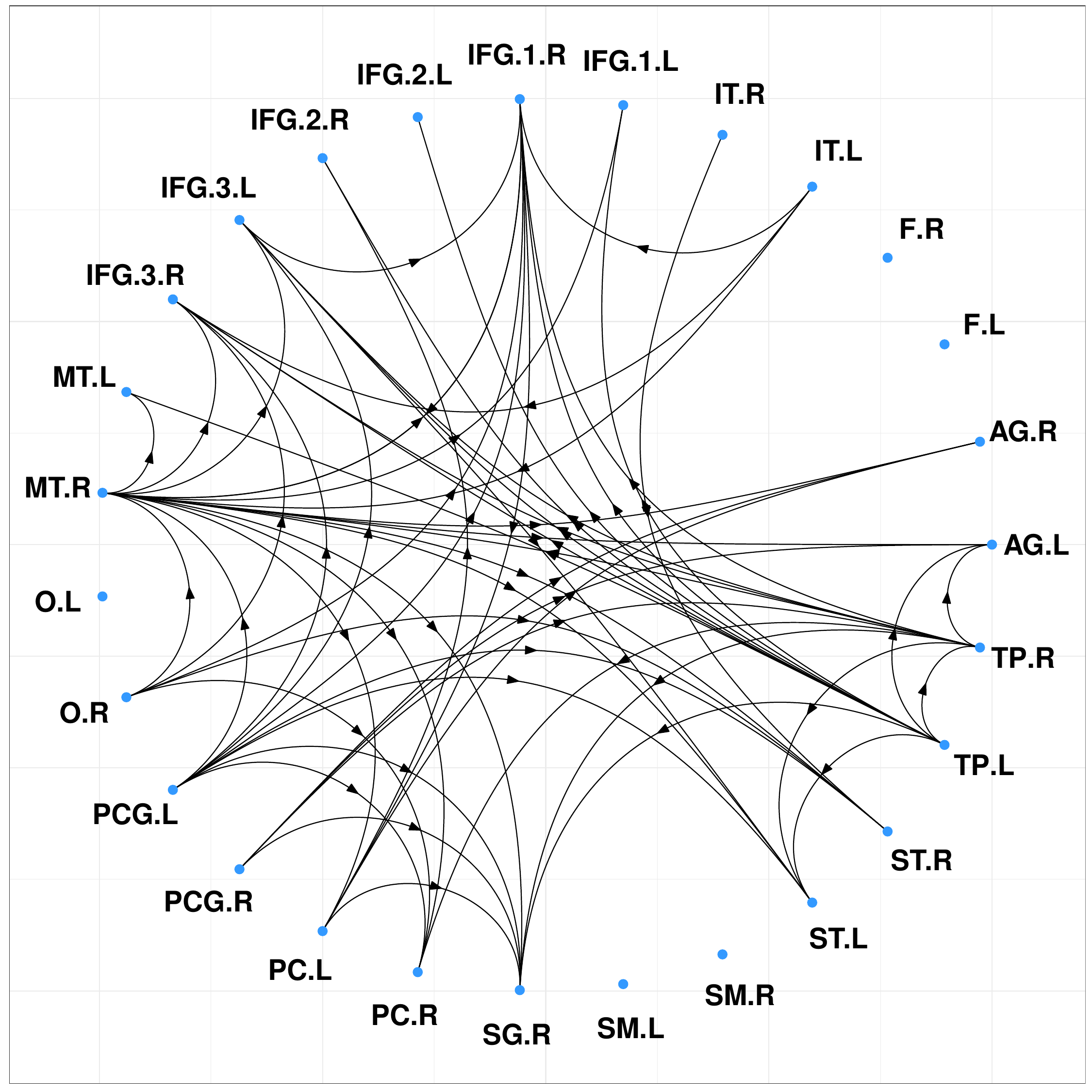}
}
\caption{Granger causality inferred from the BTVT-VAR coefficient matrix for the representative subject. \protect\subref{subfig:GC_sub4baseline_0.01_p1}
shows edges that are present in both the Granger networks at time point 1174 and 1175.\protect\subref{subfig:GC_sub4before_0.01_p1} are edges unique at time point 1174 and \protect\subref{subfig:GC_sub4after_0.01_p1} are edges unique at time point 1175.}
\label{fig:GC_sub4}
\end{figure}

The analysis of Granger causality networks enables us to create a high-resolution map of time-varying connections.
To illustrate, Figure \ref{fig:GC_sub4} shows the Granger causality networks for the representative subject, as inferred from the BTVT-VAR coefficient matrix estimated at two time points close to the focal point of the story, that is, $t=1174$ and $t=1175$. We focus on the first 100 Granger edges $(j,i)$ with the highest posterior probabilities $1-p_{j,i}(t)$. Figure \subref*{subfig:GC_sub4baseline_0.01_p1} shows edges that are present in both the Granger networks at time point 1174 and 1175.  Here the Granger network is sparse and is concentrated around two hub nodes, the right precentral gyrus (PCG.R) and both the left and right precuneus gyrus (PC.L and PC.R).  The primary motor cortex, which is located in the PCG, is usually active during any task that incorporates movement and has been associated with reading out loud \citep{smelser}.  The PC is involved in a variety of complex functions, which include recollection and memory, integration of information relating to perception of the environment, cue reactivity, mental imagery strategies, episodic memory retrieval \citep{borsook}. 
Figure \subref*{subfig:GC_sub4before_0.01_p1} and \subref*{subfig:GC_sub4after_0.01_p1} show the unique edges at time points 1174 and 1175. Hence, they reflect the brain effective connectivity dynamics.  In Figure \subref*{subfig:GC_sub4before_0.01_p1}, the right fusiform gyrus (F.R) is a hub node with many connections arising from F.R to several other ROIs.  The fusiform gyrus plays important roles in object and face recognition and the visual processing of written words.  All of the inferior frontal gyrus areas also have many directed Granger causality connections to and from several other ROIs. The left inferior frontal gyrus has a number of functions including the processing of speech and language in Broca's area \citep{kennison}. In Figure \subref*{subfig:GC_sub4after_0.01_p1}, the Granger causality network exhibits increased dispersion and density. Many regions of interest (ROIs) are linked to the left inferior frontal gyrus, opercular part (IFG.1.L), while the left middle temporal gyrus (MT.L) shows connections both from and to ROIs. The middle temporal gyrus is known for its sensitivity to visual motion. Traditional language processing areas encompass the inferior frontal gyrus (Broca’s area), superior temporal and middle temporal gyri, supramarginal gyrus, and angular gyrus (Wernicke’s area). Recent evidence suggests that structures in the medial temporal lobe also play a role in language processing \citep{tracy}.

\section{Discussion}
\label{Conclusion}
We have introduced a scalable Bayesian time-varying tensor VAR model with a few  distinctive features. The model relies on a state-space representation comprising $2^H$ elements. In this representation, each state is derived from a subset of $H$ components shared across the entire time span. The activations of these components at each time point are governed by a latent Ising model, while shrinkage priors enforce sparsity of the tensor representation. These features contribute to a more natural description of connectivity dynamics for fMRI data. Moreover, this representation aligns with the emerging belief within neuroscience that connectivity states evolve gradually, rather than undergoing abrupt changes over time as suggested by traditional models based on hidden Markov models. As a result, our model allows for a finer resolution of time-varying functional connectivity. We can estimate time-varying Granger causality networks and investigate changes in connectivity across subsequent time points. 

In our study, we demonstrate the model's efficacy and interpretability in analyzing effective connectivity in fMRI experiments. Our method extends beyond fMRI to encompass other types of time-varying neuroimaging data, including EEG data, and more generally to any data where a vector autoregressive model is appropriate. In fMRI data, where an AR(1) dependence is often assumed as sufficient \citep[however, see][for different perspectives]{Monti2011, Corbin2018}, our data analysis appears to confirm this general suggestion. Higher lags of the coefficient matrices did not exhibit patterns significantly different from zero in our fMRI experiment.

The proposed framework can be extended to obtain scalable inference and describe shared patterns of brain connectivity in multi-subject analyses,  by tracking the components active across multiple time intervals and subjects. However, this type of inference would require ensuring the identifiability of the same tensor components across subjects.  One approach to achieving this could involve utilizing clustering-inducing Bayesian nonparametric priors, facilitating the borrowing of information across all subjects in estimating the components. Due to the increased computational burden and development required, we defer its exploration to future work.

\bibliographystyle{apalike}
\bibliography{ref}

\end{document}